\documentclass[twocolumn, aps, prapplied, superscriptaddress, longbibliography]{revtex4-2}
\usepackage{float}
\usepackage{graphicx}
\usepackage{gensymb}
\usepackage{dcolumn}
\usepackage{bm}
\usepackage{textcomp}
\usepackage[intlimits]{amsmath}
\usepackage{esint}
\usepackage{braket}
\usepackage{color}
\usepackage{filecontents}
\usepackage{bbold}
\usepackage{soul} 
\usepackage{xcolor}

\begin{document}
	
	
	\title{Upper bounds on focusing light through multimode fibers}

 	\author{Amna Ammar}
	\affiliation{Institute of Materials Science and Nanotechnology, National Nanotechnology Research Center (UNAM), Bilkent University, 06800 Ankara, Turkey}
  	\author{Sarp Feykun Şener}
 \affiliation{Institute of Materials Science and Nanotechnology, National Nanotechnology Research Center (UNAM), Bilkent University, 06800 Ankara, Turkey}
 	\affiliation{Department of Physics, Bilkent University, 06800 Ankara, Turkey}
 	\author{Mert Ercan}
 \affiliation{Institute of Materials Science and Nanotechnology, National Nanotechnology Research Center (UNAM), Bilkent University, 06800 Ankara, Turkey}
 \affiliation{Department of Physics, Bilkent University, 06800 Ankara, Turkey}
	\author{Hasan Y{\i}lmaz}
        \email{hasan.yilmaz@unam.bilkent.edu.tr}
	\affiliation{Institute of Materials Science and Nanotechnology, National Nanotechnology Research Center (UNAM), Bilkent University, 06800 Ankara, Turkey}

	
\begin{abstract}

Wavefront shaping enables precise control of light propagation through multimode fibers (MMFs), facilitating diffraction-limited focusing for applications such as high-resolution single-fiber imaging and high-power fiber amplifiers. While the theoretical intensity enhancement at the focal point is dictated by the number of input degrees of freedom, practical constraints—such as phase-only modulation and experimental noise—impose significant limitations. Despite its importance, the upper bounds of enhancement under these constraints remain largely unexplored. In this work, we establish a theoretical framework to predict the fundamental limits of intensity enhancement with phase-only modulation in the presence of noise-induced phase errors, and we experimentally demonstrate wavefront shaping that approaches these limits. Our experimental results confirm an enhancement factor of 5000 in a large-core MMF, approaching the theoretical upper bound, enabled by noise-tolerant wavefront shaping. These findings provide key insights into the limits of phase-only control in MMFs, with profound implications for single-fiber imaging, optical communication, high-power broad-area fiber amplification, and beyond.
		
\end{abstract}
	
\pacs{Valid PACS appear here}
	
\maketitle
	
\section{Introduction}

Multimode fibers (MMFs) are essential for high-resolution and ultra-thin single-fiber endoscopic imaging~\cite{2012_Cizmar_NC, 2012_Choi_PRL, 2012_Papadopoulos_OE, papadopoulos2013high, loterie2015digital, amitonova2018compressive, amitonova2020endo}, optical manipulation~\cite{vcivzmar2011shaping, leite2018three}, high-bandwidth short-distance optical communication~\cite{li2021large}, precise laser-based material processing~\cite{kakkava2019selective}, and power scaling in fiber amplifiers~\cite{dawson2008analysis, richardson2010high, zervas2014high, fu2017review}. Their ability to support a large number of spatial modes increases the capacity for information transmission and energy delivery, but also introduces challenges due to modal dispersion and complex interference effects. Overcoming these challenges requires precise control over light propagation within MMFs. Wavefront shaping provides a powerful approach to manipulate interference at the fiber’s output by tailoring the input field using a spatial light modulator (SLM), enabling applications such as high-resolution imaging, targeted light delivery, and nonlinear effect management~\cite{di2011hologram, 2012_Mosk_NP_Review, 2015_Horstmeyer_NP_Review, vellekoop2015feedback, 2017_RotterR, tzang2018adaptive, gigan2022roadmap, cao2022shaping, cao2023controlling, Chen2023, wisal2024optimal}. Precise control of the output field through input wavefront shaping requires accurately measuring the fiber’s transmission matrix, which characterizes the input-output field relationship~\cite{2010_Popoff_PRL, vcivzmar2011shaping, 2012_Choi_PRL, 2015_Choi_OptExpress_R, 2021_Li_NC, bender2023spectral}. This measurement is essential for optimizing wavefront shaping techniques, including the fundamental task of focusing light at a desired location~\cite{2007_Vellekoop_OL, vellekoop2010exploiting, 2012_Papadopoulos_OE, 2013_Caravaca-Aguirre_OE, dremeau2015reference, 2016_Amitonova_OL, 2016_Descloux_OE, n2018mode, nam2020increasing, 2022_Lyu_AO, gomes2022near, lyu2024wavefront, hammer2025effect}.

A key metric for evaluating wavefront shaping performance for focusing light through complex media is the enhancement factor, which quantifies the intensity at the focal position~\cite{2007_Vellekoop_OL, 2016_Amitonova_OL}. The enhancement factor is defined as $\eta_m = I_m^{(\mathrm{foc})}/\langle I_m^{(\mathrm{rand})} \rangle$, where $I_m^{(\mathrm{foc})}$ is the focus intensity at the target position $m$ after wavefront shaping and $\langle I_m^{(\mathrm{rand})}\rangle$ is the averaged intensity at the same position $m$ over multiple independent random wavefront inputs~\cite{2007_Vellekoop_OL,  vellekoop2015feedback, mastiani2024practical}. To maximize the enhancement factor, an optimization algorithm is typically used to determine the ideal input wavefront~\cite{vellekoop2008phase, 2013_Yilmaz_BOE, dremeau2015reference, n2018mode, nam2020increasing, mastiani2021noise, mastiani2024practical}. Fundamentally, these algorithms are based on some form of transmission-matrix measurement. In principle, with complete control over both the amplitude and phase of the input channels, all transmitted power can be efficiently collected at a desired focal position~\cite{gomes2022near}. However, in practice, there are certain limitations to the focus intensity due to incomplete input channel control~\cite{2020_Resisi_APLP, mastiani2024practical}. With full-field modulation of the input wavefront, the upper bound of the enhancement factor equals the number of controlled channels, $\eta = N$. This result relies on ideal amplitude and phase control, as well as ideal mode coupling—conditions typically satisfied in short, low-loss multimode fibers where mode-dependent loss is negligible (see Appendix B). Nevertheless, most SLMs function primarily as phase-only modulators, as amplitude modulation is generally avoided in practice~\cite{2012_Mosk_NP_Review}. This is because amplitude modulation reduces the input power, making it less efficient for applications requiring high-intensity light focusing, such as broad-area fiber amplifiers and laser ablation through large-core optical fibers~\cite{kakkava2019selective, wisal2024theory, rothe2025output, 2025_Rothe_Science}. Therefore, in practical scenarios, phase-only input modulation is typically preferred; the input wavefronts can be optimally shaped to achieve constructive interference at the desired focus location at the fiber's output, maintaining a constant input power~\cite{gigan2022roadmap, cao2023controlling, vellekoop2008phase, rothe2025output}. This constraint inherently limits the maximum enhancement factor, with its theoretical upper bound, following $\eta = R(N-1)+1$, where $R = \pi/4$ represents the participation ratio—a well-established result for focusing light through disordered scattering media.~\cite{2007_Vellekoop_OL, vanBeijnum2008thesis, vellekoop2008phase, 2013_Yilmaz_BOE, vellekoop2015feedback, mastiani2024practical}. 

Although the assumption that the participation ratio is $R = \pi/4$ has been widely used to estimate the upper bound of the enhancement factor in wavefront shaping through multimode fibers (MMFs), its validity has not been critically examined. This raises a fundamental question: is the participation ratio for phase-only modulation in MMFs $R = \pi/4$ similar to that in disordered media? If so, what is the upper limit of the achievable enhancement factor, and can it be experimentally reached? 

Here, we present a comprehensive work that combines both experimental and theoretical aspects of wavefront shaping. We introduce the theoretical upper bound on the enhancement factor for phase-only modulation; moreover, we experimentally demonstrate it for focusing light through a multimode fiber (MMF). We first describe how the participation ratio $R$ depends on the basis used for phase-only modulation. Our theoretical calculations and experimental observations reveal that when wavefront modulation is performed on the Fourier space of the proximal end of the fiber, the $R_m = \pi/4$ at any fiber output position $m$, which closely resembles the well-known result observed in disordered media. In contrast, phase-only modulation in the fiber mode basis reveals a strong radial dependence of $R$ at the fiber’s distal end. To experimentally approach the theoretical upper bound of the enhancement factor for phase-only modulation, we perform noise-tolerant transmission-matrix measurements using a Hadamard basis, achieving an enhancement factor of $\eta = 5,000$. We also introduce a predictive method that combines our theoretical framework with measured phase errors in a practical wavefront shaping setup. Using this approach, we quantify the phase errors in transmission-matrix measurements both in the canonical (SLM pixel) and Hadamard bases. This method provides a clear quantitative understanding of how Hadamard-based measurements minimize phase errors, enabling near-ideal wavefront shaping. By linking theoretical predictions with experimental observations, this work establishes a foundation for accurately predicting the enhancement factor and achieving near-perfect phase-only wavefront shaping in multimode fibers. Our findings directly advance the development of more robust and efficient wavefront shaping techniques, with broad applications in high-resolution imaging, broad-area fiber amplifiers, laser ablation through multimode fibers, and beyond.

\begin{figure*}[tb]
	\centering
	\includegraphics[width=18cm]{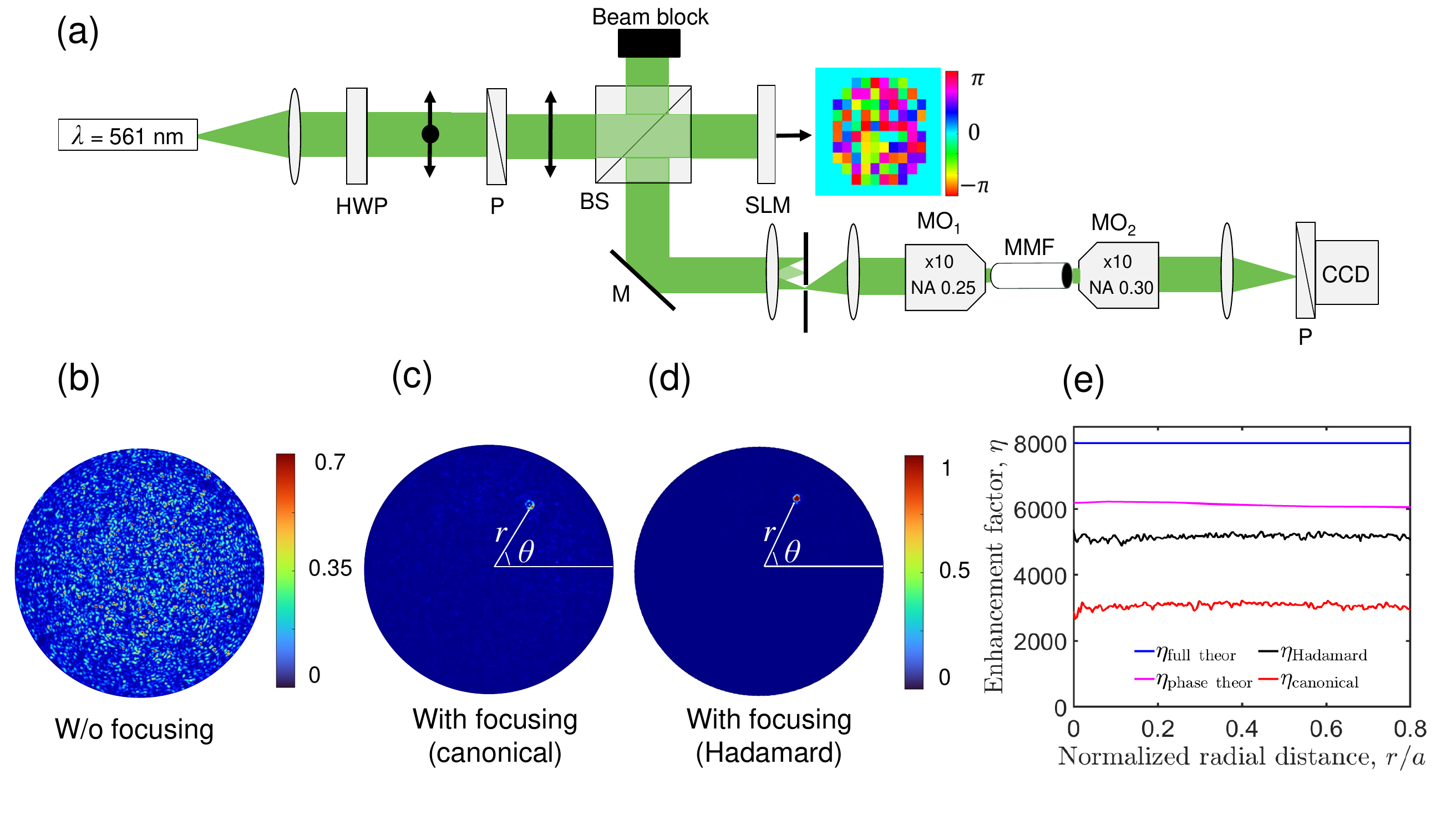}
	\caption{Wavefront shaping setup and the results are shown. (a) Experimental setup: The spatial light modulator (SLM) modulates the laser beam on the multimode fiber's proximal end and focuses on the distal end. P: linear polarizer; HWP: half-wave plate; BS: beam splitter; M: mirror; ${\rm MO}_1$ and ${\rm MO}_2$: microscope objectives; MMF: multimode fiber; NA: numerical aperture; CCD: charge-coupled device. (b) An experimental image of the speckle formation at the distal end of the MMF is shown when a random wavefront is incident on the proximal end. The interference of the waves propagating through various optical modes in the MMF results in random intensity fluctuations, giving rise to the granular appearance of speckle patterns. Here, the fiber radius is $a = 100$ \textmu m. (c) Experimental image of the distal end of the fiber when light is focused by wavefront shaping on the canonical (SLM pixel) basis and (d) on the Hadamard basis with $N = 8,000$. The scale bar indicates the intensity across the distal end as observed on the CCD camera and is normalized to the highest count on the image. (e) The mean enhancement factor $\eta$ averaged over azimuthal $\theta$ positions versus the normalized radial distance $r/a$ at the fiber distal end for the number of degrees of freedom $N = 8,000$. The blue solid line represents the upper limits of the enhancement factor with full-field modulation at the input equal to $\eta = N$. The violet solid line represents the upper bounds of the enhancement factor with perfect phase-only modulation when the SLM is placed on the Fourier plane of the fiber proximal end. The black and red solid lines represent the experimental enhancement factors with wavefront shaping on the Hadamard and canonical (SLM pixel) basis. The enhancement factor is higher with wavefront shaping on the Hadamard basis.}
	\label{figure1}
\end{figure*}

\section{Transmission-matrix and focusing measurements}

The first step in controlling light propagation through a multimode fiber (MMF) involves measuring its transmission matrix. The transmission of light through an MMF can be described with a transmission matrix with elements $t_{mn}$
\begin{align}
E_{m} = \sum_{n=1}^{N} t_{mn} E_{n}\label{eq:1}
\end{align}
where $m$ and $n$ are the indices of the outgoing and incident fields, and $N$ is the number of independently controlled degrees of freedom at the input.

Using the spatial light modulator in our experimental setup, shown in Fig.~\ref{figure1}(a), we select an input basis indexed by $n$ (either canonical or Hadamard) for the incident fields $E_n$, as defined in Equation \ref{eq:1}. The field transmission matrix is then measured using common-path phase-shifting interferometry~\cite{2010_Popoff_PRL, Wade1, Yilmaz2019, Yilmaz2019memory} in a chosen basis.

\begin{figure*}[th]
	\centering
	\includegraphics[width=18cm]{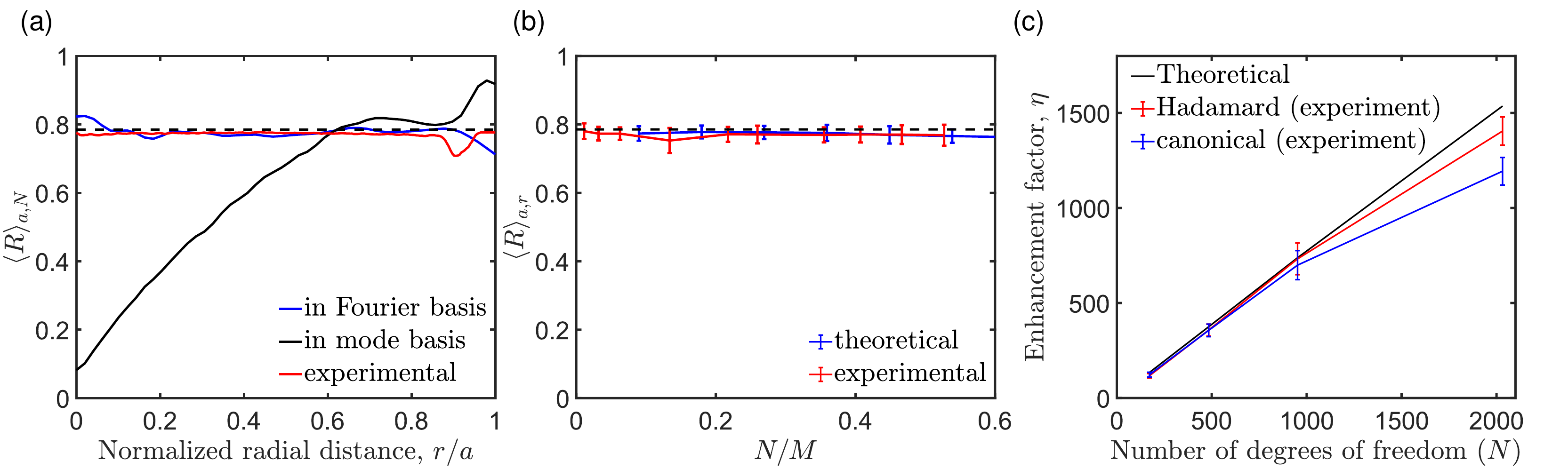}
	\caption{(a) The mean participation ratio $R$, averaged over the core radius $a$ and the number of degrees of freedom $N$, versus the normalized radial distance $r/a$ is illustrated. The experimental $R$ (solid red line) is consistent with the numerical $R$ (solid blue line) in the Fourier basis, both following a trend close to $\pi/4$ (dashed black line) and showing no dependence on the radial distance. However, we observe a strong dependence of $R$ on the radial distance when $R$ is computed with phase-only modulation on the MMF fiber mode basis (solid black line). (b) The mean participation ratio $R$, averaged over the core radius $a$ and the radial distance $r$ at the fiber distal end, is shown with respect to the normalized number of degrees of freedom $N/M$. The experimental $R$ (red line) agrees with the numerical $R$ (solid blue line) and remains invariant with $N/M$, maintaining a value near $\pi/4$ (dashed black line). (c) The mean enhancement factor $\eta$, averaged over the azimuthal position $\theta$ and radial distances $r$ at the fiber distal end, is shown with respect to the number of degrees of freedom $N$ = 172, 484, 952, and 2,032. The experimental enhancement factor $\eta$ in the Hadamard basis (red solid line) closely follows the theoretical prediction (black solid line) and is notably higher than the experimental $\eta$ in the canonical basis (blue solid line).}
	\label{figure2}
\end{figure*}

To measure the transmission matrix of the MMF using common-path interferometry~\cite{2010_Popoff_PRL}, the light field on the spatial light modulator is divided into a signal and a reference part. The signal is modulated with four phase steps uniformly spaced between $0$ and $2\pi$, and both the signal and reference components are then coupled into the MMF. The MMF is a step-index (SI) fiber with a diameter of 200 \textmu m, a length of 6 cm, and an NA = 0.22. The light propagates through the supported modes of the MMF, and the resulting intensity pattern at the distal end is imaged onto a charged-coupled device (CCD) camera sensor. In this work, we use feedback-based wavefront shaping to retrieve the transmission matrix from the SLM (in the Fourier space) to the CCD camera (in the real space) using two different algorithms: the stepwise sequential algorithm (SSA) on the canonical basis~\cite{2007_Vellekoop_OL} and the dual reference algorithm on the Hadamard basis~\cite{mastiani2021noise}. For both methods, we divide the SLM surface into  $N$ number of input degrees of freedom (segments) where $N$ is chosen to be 172, 484, 952, 2,032, 3,300, 3,940, 5,388, 6,180, 7,080, and 8,000 in all the experiments described here. We choose $N$ in such a way that we display a circular phase pattern on the SLM. In our experimental setup, depicted in Fig.~\ref{figure1}(a), we modulate a single linear horizontal polarization of light at the input and detect the same linear horizontal polarization at the output (the details are described in Appendix A).

Maximizing the overlap between the input field patterns and the fiber’s core at its proximal end, both in real and Fourier space, is essential. In real space, the input field patterns must be centered at the fiber core, and their size should be equal to or smaller than the fiber core diameter. In Fourier space, the diameter of the displayed SLM pattern should correspond to a numerical aperture equal to or smaller than the fiber’s numerical aperture, NA = 0.22. We adjusted our setup such that displaying SLM patterns with $6\times 6$ pixel-size segments ensures a near-perfect size match between any arbitrary SLM pattern and the fiber core (additional details can be found in Appendix A). Furthermore, the diameter of the SLM pattern consistently corresponds to an NA smaller than that of the fiber, with NA $<$ 0.22.
 
The phases of each transmission-matrix column is computed using four recorded CCD camera images, each corresponding to a different relative phase between the signal and the reference, obtained through common-path four-phase shifting interferometry with the SLM~\cite{2010_Popoff_PRL, Wade1, Yilmaz2019, Yilmaz2019memory}. The transmission-matrix elements map the input field onto the output field, where every element on the output field is a summation of all the input elements multiplied by an analogous transmission coefficient. Thus, a random summation of $N$ field components contributes to each element in the output field, which results in the speckle pattern as is seen in Fig.~\ref{figure1}(b). To form a focus, i.e., to increase light intensity for a specific output element, the $N$ field components must constructively interfere with each other. This is accomplished by displaying the conjugate of the measured transmission-matrix phase row with index $m$, which corresponds to the desired position of the output field, on the SLM.

In the stepwise sequential algorithm (SSA) on the canonical basis~\cite{2007_Vellekoop_OL}, the SLM is divided into $N$ segments, and four-phase shifting interferometry is applied to each segment individually. This process varies the relative phase from 0 to 2$\pi$ between the selected segment and the remaining $N - 1 $ segments, which serve as the reference signal. The procedure is repeated for all $N$ segments, allowing the measurement of the transmission-matrix elements. However, a key drawback of this method is that the signal-to-noise ratio (SNR) decreases as the number of degrees of freedom $N$ increases~\cite{2013_Yilmaz_BOE}. This occurs because the signal intensity from each segment is significantly smaller than the reference contribution from the rest of the SLM segments.

In the dual reference algorithm~\cite{mastiani2021noise}, we use the Hadamard basis, where the size of the basis must be $N_1 = 2^p$, with $p$ being an integer (e.g., $N_1$ = 128, 256, 512, 1,024, 2,048, 4,096). In this approach, we divide the SLM segments into two equal-sized groups, each containing a small number of overlapping segments, denoted as $O$. In the first step of the algorithm, a Hadamard pattern is displayed on the segments of group 1 (segments from 1 to $N_1$), while the remaining segments (from $N_1 + 1$ to $N$) contribute as the reference field $E_{\rm ref}^{(1)}$. Four-phase shift interferometry is applied to both groups to vary their relative phase from 0 to $2\pi$~\cite{2010_Popoff_PRL}. Subsequently, we perform a Hadamard transform on the full-field patterns obtained from the four-phase-shifting interferometry to extract the transmission-matrix phase elements for group 1 relative to the $E_{\rm ref}^{(1)}$ phase, which results in the transmission matrix 1 $t^{(1)}$.

In the second step of the algorithm, the segments of group 2 (from $N-N_1+1$ to $N$) are modulated, while the remaining segments (from 1 to $N-N_1$) serve as reference $E_{\rm ref}^{(2)}$. The procedure is then repeated, producing two matrices, $t^{(1)}$ and $t^{(2)}$, with phase values relative to the two different reference fields $E_{\rm ref}^{(1)}$ and $E_{\rm ref}^{(2)}$ for each camera pixel $m$. To obtain the final transmission matrix, we need to determine the phase difference between $E_{\rm ref}^{(1)}$ and $E_{\rm ref}^{(2)}$ at each camera pixel $m$. By using the phase differences between the overlapping segments $O$ relative to 
$E_{\rm ref}^{(1)}$ and $E_{\rm ref}^{(2)}$ in both steps of the algorithm, we calculate the phase difference between $E_{\rm ref}^{(1)}$ and $E_{\rm ref}^{(2)}$. Finally, we adjust the phases of $t^{(1)}$ and $t^{(2)}$ based on this known phase difference and combine the two matrices to construct the final full-field transmission matrix.

Note that, in all transmission-matrix experiments, we recorded fiber output intensity images $I_m^{(\mathrm{rand})}$ for 1,000 independent random wavefront inputs, using the same number of segments $N$ as in each experiment. Fig.~\ref{figure1}(b) shows an example random speckle pattern observed on the CCD camera for a random input wavefront. These speckles result from modal dispersion, where each fiber mode propagates with a distinct propagation constant and phase delay, leading to complex interference at the output. 

Immediately after each transmission-matrix measurement, we perform focusing experiments. To focus on a specific position $m$  within the core at the fiber’s distal end, we display the conjugated phase from the measured transmission-matrix row corresponding to position $m$ and record the focused intensity pattern on the CCD camera. To avoid CCD camera saturation, calibrated neutral density (ND) filters were placed in front of the camera during the measurements. These experiments are performed to focus on various radial $(r)$ and azimuthal $(\theta)$ positions within the core of the fiber's distal end using phase-only input modulation, obtained from the corresponding transmission-matrix data. The resulting enhancement factors are calculated using the expression $\eta_m = I_m^{(\mathrm{foc})}/\langle I_m^{(\mathrm{rand})} \rangle$, where $I_m^{(\mathrm{foc})}$ represents the focused intensity at the target position $m$ after wavefront shaping, and $\langle I_m^{(\mathrm{rand})}\rangle$ represents the averaged intensity at the same position $m$ over 1,000 independent random wavefront inputs. Two example intensity patterns are shown in Fig.~\ref{figure1}(c) and Fig.~\ref{figure1}(d) for wavefront shaping on the canonical and Hadamard basis, respectively. As evident from the captured images, the optimized focus achieved using the Hadamard basis exhibits a higher peak intensity compared to the focus obtained on the canonical basis.

Theoretically, for full-field modulation, the expected enhancement factor is given by $\eta = N$, where $N$ represents the number of input degrees of freedom. Thus, for $N = 8,000$, the theoretical full-field enhancement is $\eta = 8,000$. Our numerical simulations confirm that the expected prefactor for phase-only modulation is $\pi/4$, leading to a theoretical enhancement factor of $\eta = 6,135$ for $N = 8,000$.

Fig.~\ref{figure1}(e) presents the enhancement factor $\eta$ as a function of the normalized radial distance $r/a$, comparing wavefront shaping using the transmission matrix measured in the canonical and Hadamard bases. The results clearly show that using the Hadamard basis yields a higher enhancement factor across all radial distances within the core at the fiber's distal end. 

For imaging applications, not only the enhancement factor but also the spatial shape of the focus, i.e., the point spread function (PSF), is important. While phase-only wavefront shaping in the Fourier plane at the proximal end of the fiber can yield focal spots with a uniform enhancement factor $\eta$ across the core at the distal end, as shown in Fig.~\ref{figure1}(e), this does not imply a position-invariant PSF within the core~\cite{2016_Descloux_OE}.

\begin{figure*}[th] 
	\centering
	\includegraphics[width=18cm]{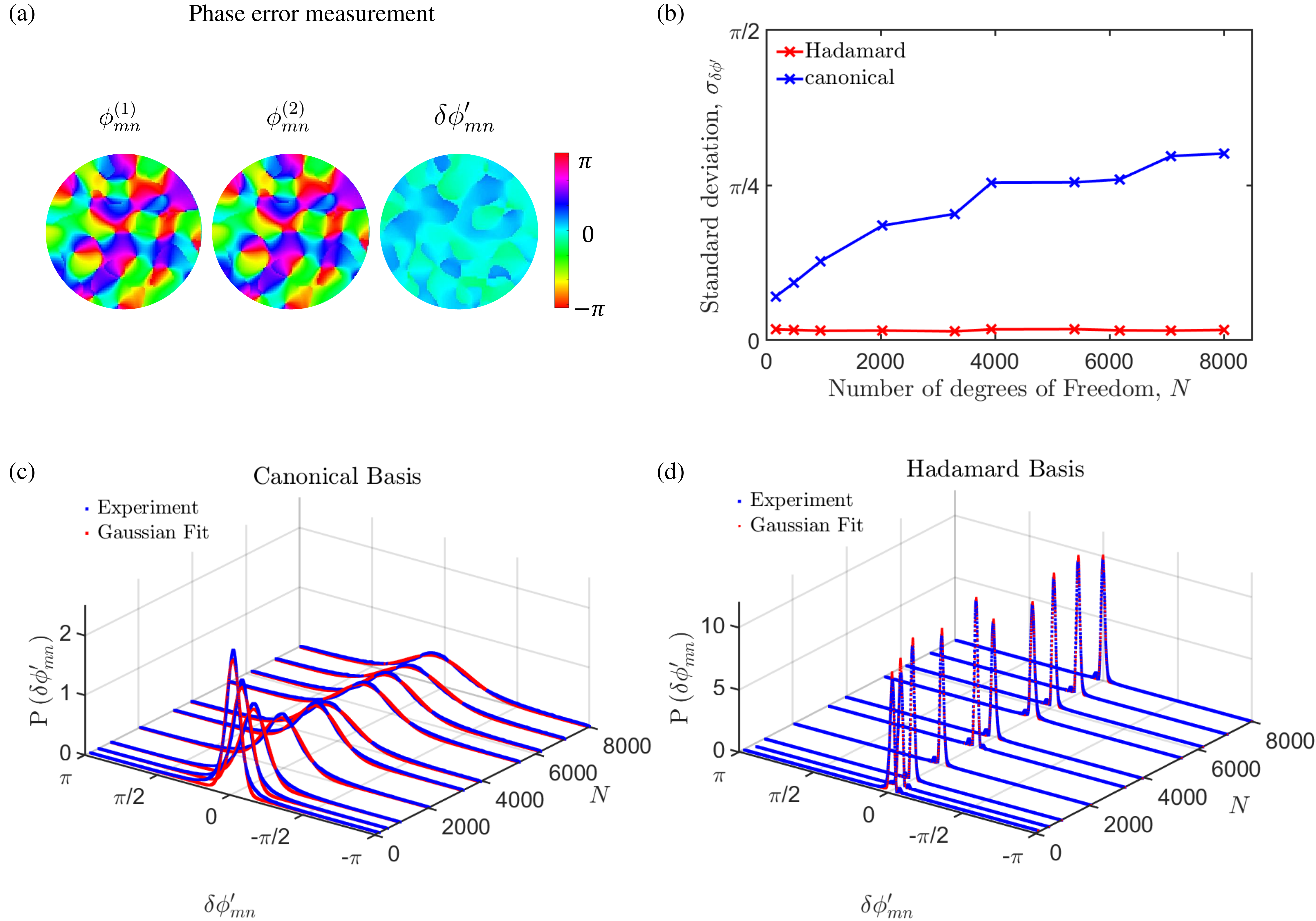}
	\caption{Phase-error measurements in wavefront shaping experiments are shown. (a) A conceptual sketch of two independently measured phase maps, $  \phi_{mn}^{(1)} $ and $ \phi_{mn}^{(2)} $ for the same input $n$. The phase difference, $\delta \phi_{mn}' = \arg \left(e^{i(\phi_{mn}^{(2)} - \phi_{mn}^{(1)})} \right)$, represents the measured phase error. (b) Standard deviation $ \sigma\mathrm{_{\delta \phi^\prime}}$ of the Gaussian-fitted phase-error distributions as a function of the number of degrees of freedom $ N $. In the canonical basis, $ \sigma\mathrm{_{\delta \phi^\prime}}$ increases with $ N $, indicating higher phase errors. In contrast, the Hadamard basis maintains a consistently low $ \sigma\mathrm{_{\delta \phi^\prime}}$, suggesting greater robustness to phase errors. (c, d) Probability density functions of the phase errors $P(\delta \phi_{mn}')$ for the canonical and Hadamard bases, respectively, as a function of $ N $. The Gaussian-fitted curves (red) show that in the canonical basis, the phase-error distribution broadens significantly with increasing $ N $. In contrast, (d) shows that the Hadamard basis maintains a sharply peaked distribution around zero, indicating minimal phase errors.}
	\label{figure3}
\end{figure*}

\begin{figure*}[th] 
	\centering
	\includegraphics[width=\linewidth]{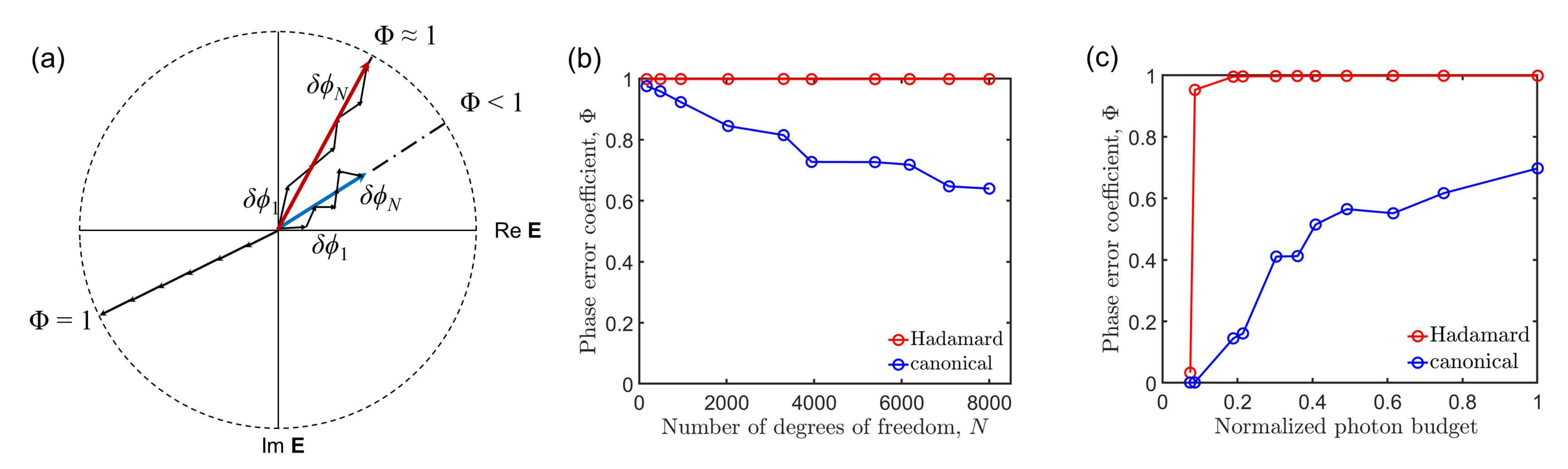}
	\caption{The impact of phase errors on the enhancement factor is illustrated. (a)  Schematic representation of the phase-error coefficient $\Phi$ for the Hadamard (red) and canonical (blue) bases, where $\delta\phi$ represents phase errors. The dashed lines indicate how phase errors accumulate differently in the two bases, with the Hadamard basis maintaining $\Phi \approx 1$ due to a reduced cumulative effect of phase errors.
    (b) The phase-error coefficient $\Phi$ as a function of the number of degrees of freedom $N$ is shown. On the canonical basis, $\Phi$ decreases with $N$ due to a decreasing signal-to-noise ratio (SNR), which reduces interferometric visibility. In contrast, the Hadamard basis maintains $\Phi \approx 1$ due to a balanced signal-to-reference ratio. (c) The phase-error coefficient $\Phi$ for $N = 8000$, plotted as a function of the normalized photon budget. The Hadamard basis maintains a consistently high $\Phi$ compared to the canonical basis, reflecting its improved SNR under the same noise conditions. The normalized photon budget (ranging from 0 to 1) is defined so that a value of 1 corresponds to a maximum photon budget of approximately 214 mean counts on the CCD.
    }
	\label{figure4}
\end{figure*}

\section{The enhancement factor and its practical limitations}

We present a general form of the enhancement factor at the output $m$ in terms of two major parameters: the input participation ratio $R$ and the phase-error coefficient $\Phi$:
\begin{align}
\eta_m = \alpha R_m\Phi_m(N-1) +1,
\end{align}
where $\alpha \equiv \left \langle A_n  \right \rangle_n^2/\left \langle A_n^2  \right \rangle_n$ defines the spatial homogeneity of the amplitude profile incident on the SLM, $R_m \equiv \left \langle |t_{mn}|  \right \rangle_n^2/\left \langle |t_{mn}|^2  \right \rangle_n$ quantifies the input participation ratio at the output $m$, $\Phi_m \equiv \left \langle {\rm cos}(\delta \phi_{mn}) \right \rangle_n^2$ quantifies the phase-error coefficient at the output $m$, and $N$ represents the number of controlled degrees of freedom at the input wavefront. Here $\left\langle\right\rangle_n$ denotes averaging over inputs $n$. The phase error coefficient $\Phi$ varies between 0 and 1, where, in the absence of phase errors, it equals $\Phi = 1$~\cite{vanPutten2011thesis}. In our experiments, $\alpha = 1$, since we illuminate the SLM with a flat-top laser beam expanded from a single-mode fiber output. In our setup, each input degree of freedom, indexed by $n$, corresponds to a $6\times 6$ pixel segment on the SLM (approximately representing a single Fourier component at the proximal end), while each output index $m$ corresponds to a CCD camera pixel, imaging the distal end.

The enhancement factor used in this work is defined as the ratio between the intensity at the target position $m$ after wavefront shaping and the average intensity at the same position $m$ measured over many independent random input wavefronts. In some studies, however, the enhancement factor is instead defined as the ratio between the focused intensity and the average background intensity after focusing—an approach that effectively serves as a proxy for the signal-to-noise ratio (SNR)~\cite{gomes2022near, duan2023modulate, 2010_Popoff_PRL}.

When the number of controlled degrees of freedom $N$ is much smaller than the total number of modes $M$ ($N \ll M$), these two definitions are equivalent. As $N$ increases, however, non-Gaussian contributions to the intensity correlations become significant. Long-range correlations, for instance, can raise the background intensity level, as has been experimentally demonstrated for wavefront shaping and focusing through disordered media~\cite{2008_Vellekoop_PRL, shaughnessy2024multiregion}. This makes the two definitions diverge in the non-Gaussian regime.

Multimode fibers (MMFs) represent a distinct case—particularly when nearly all propagating modes can be controlled ($N \approx M$) and the transmission matrix is nearly unitary. In the ideal limit of full-field (amplitude and phase) control, focusing is exact, and the SNR diverges as $N \to M$, being limited only by practical constraints rather than theoretical bounds.

In this work, we adopt the standardized definition of the enhancement factor~\cite{2007_Vellekoop_OL, 2008_Vellekoop_PRL, vellekoop2008phase, 2012_Papadopoulos_OE, 2013_Yilmaz_BOE, vellekoop2015feedback, 2016_Amitonova_OL, 2020_Resisi_APLP, mastiani2024practical}, which has a well-defined upper bound of $\eta = M$ for complete ($N = M$) full-field input modulation and $\eta = R(N-1)+1$ for phase-only input modulation (see Appendix B for details).

\subsection{Enhancement factor for phase-only input modulation}

In phase-only wavefront shaping experiments, evaluating the input participation ratio $R_m$ is essential, as it quantifies the fraction of input degrees of freedom that effectively contribute to the output at position $m$ under phase-only modulation. We quantify $R_m$ both in experiments and numerical simulations. To obtain the numerical participation ratio, we first compute the transmission matrix using the mode decomposition method~\cite{Lee2023GitHub} (see Appendix B for details). To evaluate the participation ratio in the desired input basis, we apply a basis transformation to the columns of the transmission matrix and compute the participation ratio in the transformed basis. To obtain the experimental input participation ratio, we complemented phase measurements of the transmission matrix—acquired using the dual reference algorithm in the Hadamard basis—with amplitude measurements by coupling only the signal component of the light from the SLM into the fiber. This approach allowed us to reconstruct both the amplitude and phase of the transmission-matrix elements $t_{mn'}$, where $n'$ represents the Hadamard vector index. We then applied a Hadamard transform as $t_{mn} = \sum_{n'=1}^{N} t_{mn'}H_{n'n}$ to convert the transmission matrix into the canonical (SLM pixel) basis $n$. Here $H_{n'n}$ represents the unitary Hadamard transform matrix. The participation ratio was subsequently calculated on this basis, as our phase-only modulation experiments are performed in canonical bases.

Fig.~\ref{figure2}(a) presents both numerical and experimental participation ratios $R$ as a function of the normalized radial distance $r/a$ within the core at the distal end. The numerical results in the Fourier basis closely align with the experimental data, exhibiting the same overall trend. Notably, when phase-only input modulation is performed on the Fourier basis, $R$ remains constant regardless of the focal position at the distal end of the fiber. Additionally, placing the spatial light modulator (SLM) on the Fourier plane of the MMF's proximal end yields a participation ratio of $R = \pi/4$, consistent with the well-known value observed in phase-only wavefront shaping through disordered media~\cite{vanBeijnum2008thesis}. In the experiments, the number of guided modes is $M = 15,178$ for a fiber with a core radius of $a = 100$ \textmu m and numerical aperture (NA) of 0.22. In the simulations, we consider four fibers with core radii of $a$ = 15 \textmu m, 22 \textmu m, 28 \textmu m, and 38 \textmu m (all with NA = 0.22), corresponding to $M$ = 180, 374, 606, and 1,114, respectively. All numerical values are calculated for light with a wavelength of 561 nm.

While implementing phase-only modulation directly on the fiber mode basis is not practical, it offers valuable insights into the fundamental behavior of phase-only wavefront shaping. For this reason, we also calculated the participation ratio numerically on a fiber mode basis. Our simulations reveal that when phase-only modulation is applied on this basis, the participation ratio $R$ shows a strong dependence on the radial distance of the focal point at the distal end of the MMF. At the center of the fiber core, the participation ratio drops below 0.2, as most fiber mode wavefunctions exhibit ring-shaped profiles and contribute minimally at the core center.

Fig.~\ref{figure2}(b) shows the radially averaged theoretical and experimental participation ratios $R$ (both for Fourier basis) as a function of the normalized number of degrees of freedom $N/M$, where the experimental $M$ is determined from the equation $V = k_0a\text{NA}$, with $k_0 = 2\pi/\lambda$, $\lambda = 561$ nm, $a = 100$ \micro m (core radius), and NA = 0.22. For our fiber, the number of modes per polarization is $M = V^2/4 = 15,178$. Our experimental and numerical results consistently show that when phase-only input modulation is performed on the Fourier basis, the participation ratio remains $R = \pi/4$, regardless of the number of input degrees of freedom.

Fig.~\ref{figure2}(c) presents the enhancement factor $\eta$, averaged over radial ($r$) and azimuthal ($\theta$) positions, as a function of $N$. We observe that the experimental enhancement factor closely approaches the theoretical limit without phase errors, $\eta = (\pi/4)(N-1)+1$, when the transmission matrix is measured on the Hadamard basis. For $N = 2,000$, the enhancement factor obtained in the canonical basis is lower than the theoretical prediction, as the signal-to-noise ratio decreases with increasing $N$, leading to phase errors ($\Phi < 1$).

Our theoretical and experimental results show that the widely recognized $\pi/4$ factor in wavefront shaping through disordered media also appears in step-index multimode fibers when phase-only modulation is applied at the Fourier plane of the fiber’s proximal end. Numerical simulations further indicate that this $\pi/4$ factor persists for graded-index multimode fibers under the same modulation scheme, except near the core edge (see Appendix B). This correspondence arises because each Fourier-space degree of freedom couples randomly to multiple superpositions of fiber modes, yielding output speckle statistics that closely resemble those of disordered media.

Our framework for predicting the enhancement factor $\eta$ via the participation ratio $R$ is general and applies to any linear optical medium for which a transmission matrix can be obtained. Whether the medium is a scattering slab, a chaotic cavity, a step-index MMF, or a graded-index MMF, the theory holds as long as the transmission matrix accurately describes the input--output field relationship (see Appendix B for details).

\subsection{Effect of phase errors on enhancement factor}

Accurate transmission-matrix measurements are crucial for approaching the theoretical upper bound of the enhancement factor. However, experimental noise introduces phase errors that degrade these measurements~\cite{vellekoop2008phase, Yilmaz2019}. Here, we introduce a predictive methodology to quantify these phase errors by applying the phase-shift interferometry measurement twice for the same SLM input pattern. In the absence of phase errors, the measured phases $\phi_{mn}^{(1)}$ and $\phi_{mn}^{(2)}$ must be identical for the same input $n$ and output $m$. However, in practice, experimental noise gives rise to phase errors, $\delta\phi_{mn}^{(1)}$ and $\delta\phi_{mn}^{(2)}$, in both measurements. We extract these phase errors as $\delta \phi_{mn}' = \arg \left(e^{i(\phi_{mn}^{(2)} - \phi_{mn}^{(1)})} \right)$. A sketch is shown in Fig.~\ref{figure3}(a) to describe the phase-error measurement concept. Assuming the phase errors follow a Gaussian distribution with zero mean—which is a reasonable assumption as evidenced by Figs.~\ref{figure3}(c) and Figs.~\ref{figure3}(d)—we estimate the phase errors $\delta\phi_{mn}$ in a single four-phase shift interferometry measurement (see Appendix A for details).

Fig.~\ref{figure3}(b) presents the standard deviation $\sigma_{\delta\phi^\prime}$ of the least-squares fitted Gaussian functions to the experimental phase-error histograms for different numbers of degrees of freedom $N$. The Hadamard basis maintains a consistently low $\sigma_{\delta\phi^\prime}$, indicating that phase errors remain minimal variations. In contrast, the canonical basis exhibits an increasing $\sigma_{\delta\phi^\prime}$ as $N$ increases, revealing a broadening phase-error distribution. This trend suggests that phase errors in the canonical basis become more pronounced as the number of degrees of freedom increases, leading to greater deviations from ideal constructive interference.

Wavefront shaping relies on precise constructive interference at the target position $m$, which makes it sensitive to experimental noise that introduces phase errors. To characterize the level of constructive interference in our experiments, we introduce the phase-error coefficient $\Phi_m \equiv \left \langle {\rm cos}(\delta \phi_{mn}) \right \rangle_n^2$ at the target position $m$. This coefficient equals unity in the absence of phase errors and approaches zero when the phase measurements contain no usable information. Fig.~\ref{figure4}(a) illustrates how phase deviations $\delta\phi_{mn}$ disrupt the alignment of field contributions, reducing $ \Phi $ at the output position $m$. In an ideal wavefront shaping experiment, the phases of the individual field contributions are perfectly aligned, leading to strong constructive interference and high intensity at the target. In practical, high-quality experiments, small phase errors reduce this alignment slightly, resulting in $\Phi$ values close to 1. However, in canonical basis measurements, $\Phi$ is generally much lower—especially for large $N$—due to more significant phase mismatches.

To quantify the impact of accumulated phase errors, we experimentally quantified the phase-error coefficient $\Phi$ at each output position $m$ and computed its ensemble-averaged value $\langle \Phi_{mn}\rangle_n$ as a function of the number of degrees of freedom $N$, as shown in Fig.~\ref{figure4}(b). In the Hadamard basis, $\Phi$ remains close to 1, while in the canonical basis, $\Phi$ decreases significantly as $N$ increases. Hadamard-based dual-reference transmission-matrix measurement maintains a high signal-to-noise ratio by balancing the signal and reference fields, ensuring maximum interferometric visibility. In contrast, on the canonical basis, as $N$ increases, the signal diminishes, reducing the signal-to-noise ratio, degrading the visibility of interference, and limiting the accuracy of the measured transmission matrix.

Phase singularities—points of zero intensity in the reference speckle patterns that lead to undefined phases—can, in principle, compromise the accuracy of common-path interferometry–based transmission-matrix measurements. In our setup, speckles are sampled well above the Nyquist criterion (more than three CCD pixels per speckle grain), ensuring that all pixels receive non-zero intensity and effectively mitigating this issue. Moreover, our Hadamard-based dual-reference measurement scheme balances the signal and reference, further enhancing the robustness of phase retrieval. As a result, null phases do not degrade the measured enhancement factor. For comparison, in sequential stepwise algorithms, phase singularities can introduce phase errors; however, such effects are inherently captured by the phase-error coefficient $\Phi$, which quantifies all sources of phase inaccuracy.

To further explore how the photon budget influences phase errors, we examine the behavior of $\Phi$ for a fixed $N = 8000$ as a function of the incident power on the CCD camera (seen in Fig.~\ref{figure4}(c)). The incident power is adjusted by placing neutral density (ND) filters in front of the CCD chip. The Hadamard basis maintains $\Phi \approx 1$ even as the available photon budget decreases, demonstrating its robustness to photon-limited conditions~\cite{jang2017optical}. This stability arises because Hadamard-based measurements preserve high interferometric contrast and balanced signal-to-reference ratios, maintaining a reliable signal-to-noise ratio for phase retrieval. However, on the canonical basis, $\Phi$ steadily drops as the photon budget decreases. The degradation stems from a reduced signal-to-noise ratio, leading to increased phase errors. As expected, when the signal photon count approaches zero, phase-error coefficient estimation deteriorates for both bases, ultimately preventing phase retrieval.

\section{Results and Discussion}

 \begin{figure}[ht]
	\centering
	\includegraphics[width=\linewidth]{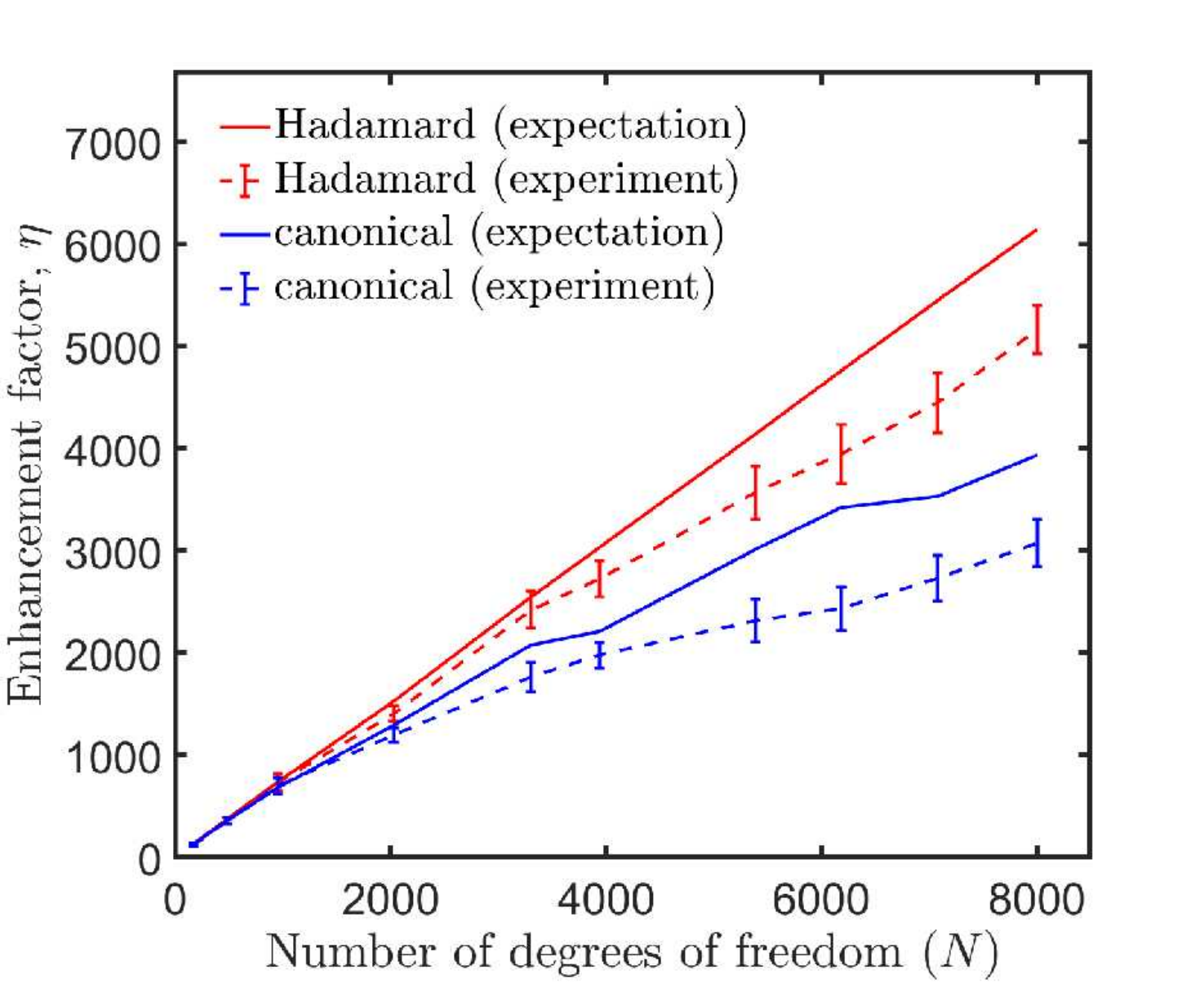}
	\caption{The mean enhancement factor $\eta$, averaged over the azimuthal $\theta$ position and radial distance $r$ at the fiber's distal end, as a function of the number of degrees of freedom $N$ is shown. Solid blue and red lines show the theoretical enhancement factors, while dashed lines represent experimental measurements for wavefront shaping on the canonical and Hadamard bases, respectively. Error bars indicate the standard deviation of enhancement factors measured at different focal positions on the fiber’s distal end.}
	\label{figure5}
\end{figure}

Here, we report the enhancement factor values for focusing at various radial distances ($r$) and azimuthal angles ($\theta$) within the fiber core at the distal end, for different values of $N$, using transmission matrices measured in both the canonical and Hadamard bases. Additionally, we provide the predicted enhancement factors obtained from our method, which uses the measured phase-error coefficients $\Phi$ for both bases.

In Fig.~\ref{figure5}, we display the measured and predicted enhancement factor $\eta$ as a function of the number of degrees of freedom, $N$. Transmission-matrix measurements in the Hadamard basis result in improved enhancement factors, thanks to its superior signal-to-noise ratio, particularly as $N$ increases. However, for $N < 952$, the results from both the canonical and Hadamard bases are nearly identical, indicating that the advantages of the dual-reference method in the Hadamard basis become more pronounced as the number of degrees of freedom grows.

For lower values of $N$ ($N < 3000$), the measured enhancement factor closely aligns with the expected theoretical values. For larger $N$, however, the measured enhancement deviates from the theoretical predictions, consistently remaining lower due to experimental limitations. Below $N = 3{,}000$, phase errors caused by experimental noise are the primary limiting factor. Beyond this threshold ($N > 3{,}000$), transmission-matrix decorrelation becomes the dominant constraint, especially during measurement periods extending up to four hours.

\section{Conclusions}

In conclusion, we present a methodology to predict the upper bound of the enhancement factor for focusing light through a multimode fiber using phase-only modulation. Our approach combines theoretical analysis with experimental phase-error measurements obtained during the acquisition of the transmission matrix. Using phase-only modulation of the incident wavefront, we experimentally approach this theoretical upper limit. The theoretical model we present, which relates the enhancement factor $\eta$ to the participation ratio $R$ and the phase-error coefficient $\Phi$, does not rely on assumptions specific to any particular class of complex media—such as scattering media, step-index fibers, graded-index fibers, or chaotic cavities. Rather, it applies to any linear optical system for which the transmission matrix can be accurately measured. Our theoretical and experimental results demonstrate that the widely recognized $\pi/4$ factor in wavefront shaping through disordered media also arises in multimode fibers when phase-only modulation is applied at the Fourier plane of the fiber’s proximal end. This coincidence arises because the field associated with each SLM segment (Fourier component) couples randomly to multiple superpositions of fiber modes, yielding output speckle statistics that are statistically equivalent to those of disordered media.

Furthermore, our method provides a quantitative explanation for the differences in enhancement factors observed in wavefront shaping experiments using Hadamard and canonical bases, attributing them to phase errors. This insight opens up possibilities for optimizing wavefront shaping techniques. Our methodology not only predicts the upper limit of the enhancement factor for focusing light through complex media but also shows how to approach it experimentally, establishing a performance benchmark grounded in physical constraints. Reaching this fundamental limit demonstrates that the theoretical bound is attainable in practice. This benchmark enables the quantitative evaluation of wavefront shaping experiments, clarifying whether performance limitations arise from fundamental constraints or implementation imperfections—an essential step toward robust, high-fidelity applications, particularly in fiber-based imaging, amplifiers, and communication systems. The upper limit we establish is especially valuable for phase-only modulation applications, such as laser ablation through large-core optical fibers~\cite{kakkava2019selective}, nonlinear effect suppression, and clean beam formation for broad-area fiber amplifiers~\cite{Chen2023, wisal2024optimal, wisal2024theory, wisal2024theory_PRX, rothe2025output, 2025_Rothe_Science}.

\begin{acknowledgements}
We thank Mesut Laçin and F. Ömer İlday for their technical support for fiber preparation and A. Serhan Başdemirci for useful discussions. This work is financially supported by the TÜBİTAK grant no. 122F311 and by the BAGEP Award of the Science Academy, with funding supplied by Sevinç - Erdal İnönü Vakfı.
\end{acknowledgements}

\section*{Appendix A: Experimental Setup}

\begin{figure*}[th]
	\centering
	\includegraphics[width=16cm]{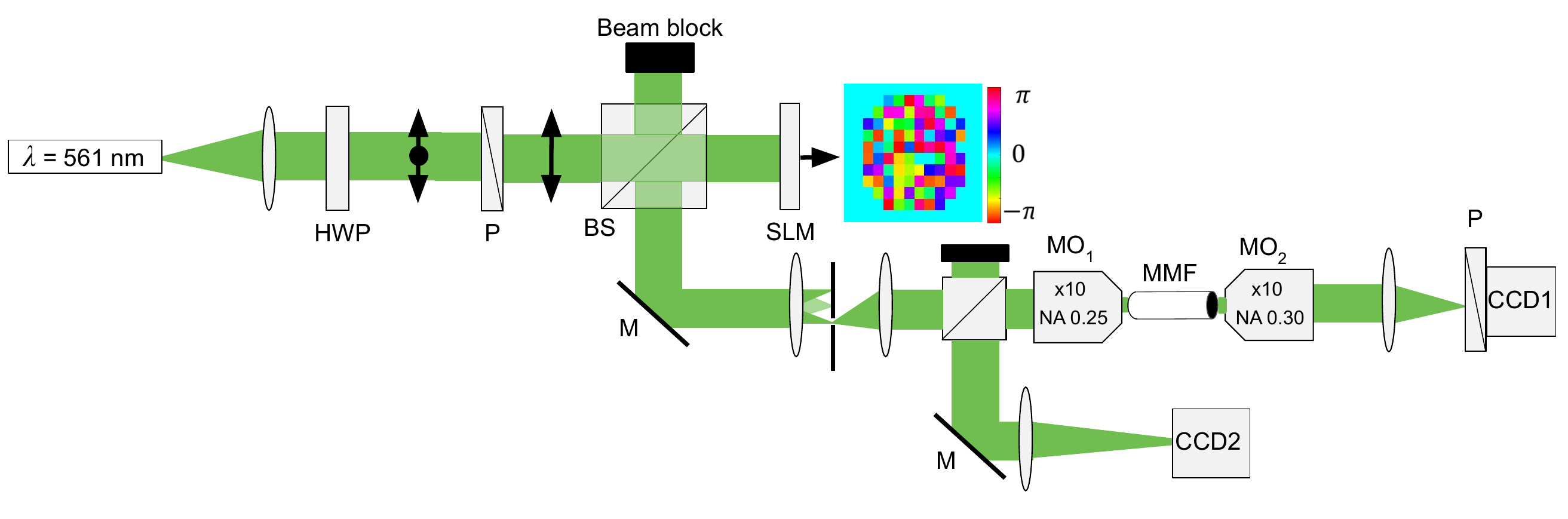}
	\caption{The experimental setup performs wavefront shaping at the distal end of the multimode fiber (MMF) and reflection measurements at its proximal end. A reflective phase-only spatial light modulator (SLM) modulates the phase front of a monochromatic laser beam ($\lambda = 561$ nm). The field transmission matrix of the multimode fiber is measured with the SLM and the CCD1 camera. The CCD2 camera is used to measure the reflected intensity patterns for aligning the incident wavefronts with respect to the fiber core at the proximal end.}
	\label{figure6}
\end{figure*}

\begin{figure*}[tb]
	\centering
	\includegraphics[width=15cm]{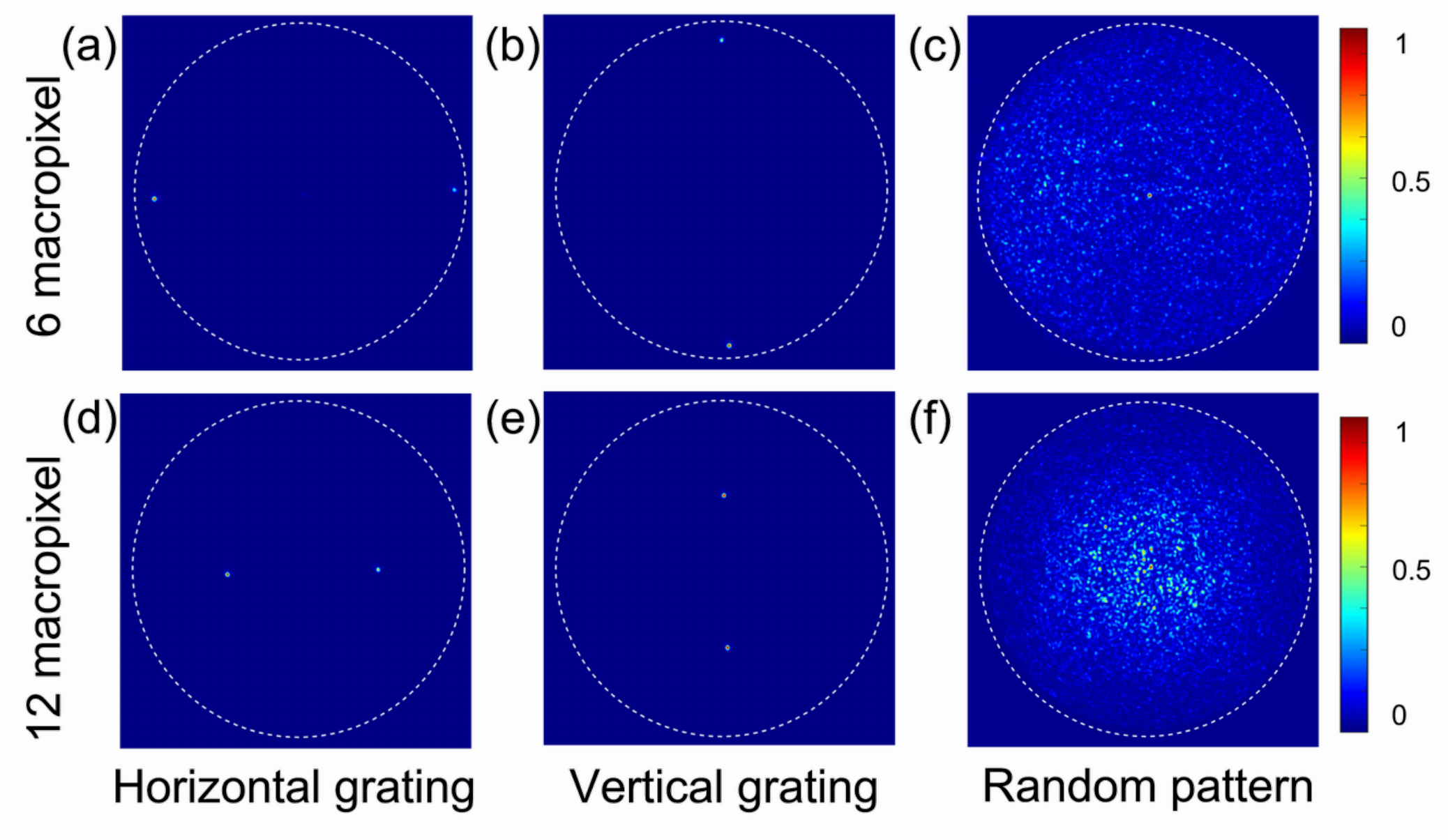}
	\caption{ Experimental images of the reflected light intensity patterns at the proximal end of the MMF are shown. (a), (b) and (c) show the reflection patterns when horizontal grating, vertical grating, and random phase patterns are displayed on the SLM using $6 \times 6$ pixel-sized segments, respectively. (d), (e), and (f) display the corresponding reflected intensity patterns for $12 \times 12$ pixel-sized segments.}
	\label{figure7}
\end{figure*}

In our experimental setup, shown in Fig.~\ref{figure6}, we use a continuous-wave (CW) laser (Coherent OBIS LS, $\lambda = 561$ nm, 120 mW, fiber-coupled with FC connector) as the light source. The laser beam is collimated by a lens and then passes through a half-wave plate (HWP) and a linear polarizer (P) to ensure that only the horizontal polarization state transmits. A beam splitter (BS) directs part of the beam toward a phase-only spatial light modulator (SLM, Meadowlark Optics, $1920 \times 1152$ pixels), which modulates the wavefront before it is coupled into the multimode fiber (MMF). The SLM operates with a phase modulation range of $0$ to $2\pi$ in discrete steps of $2\pi$/160, where 160 corresponds to the device’s dynamic range. To ensure the unmodulated, zeroth-order reflected light is not coupled into the fiber, a binary phase $(0, \pi)$ diffraction grating with a period of 6 pixels is displayed on the SLM, and only the first-order diffraction is directed to the back aperture of the microscope objective MO$_1$. Segments consist of $6 \times 6$ SLM pixels superimposed on the displayed binary phase diffraction grating. The displayed binary phase grating shifts the modulated light from the segments into the first diffraction order. 

\begin{figure}[h]
	\centering
	\includegraphics[width=9cm]{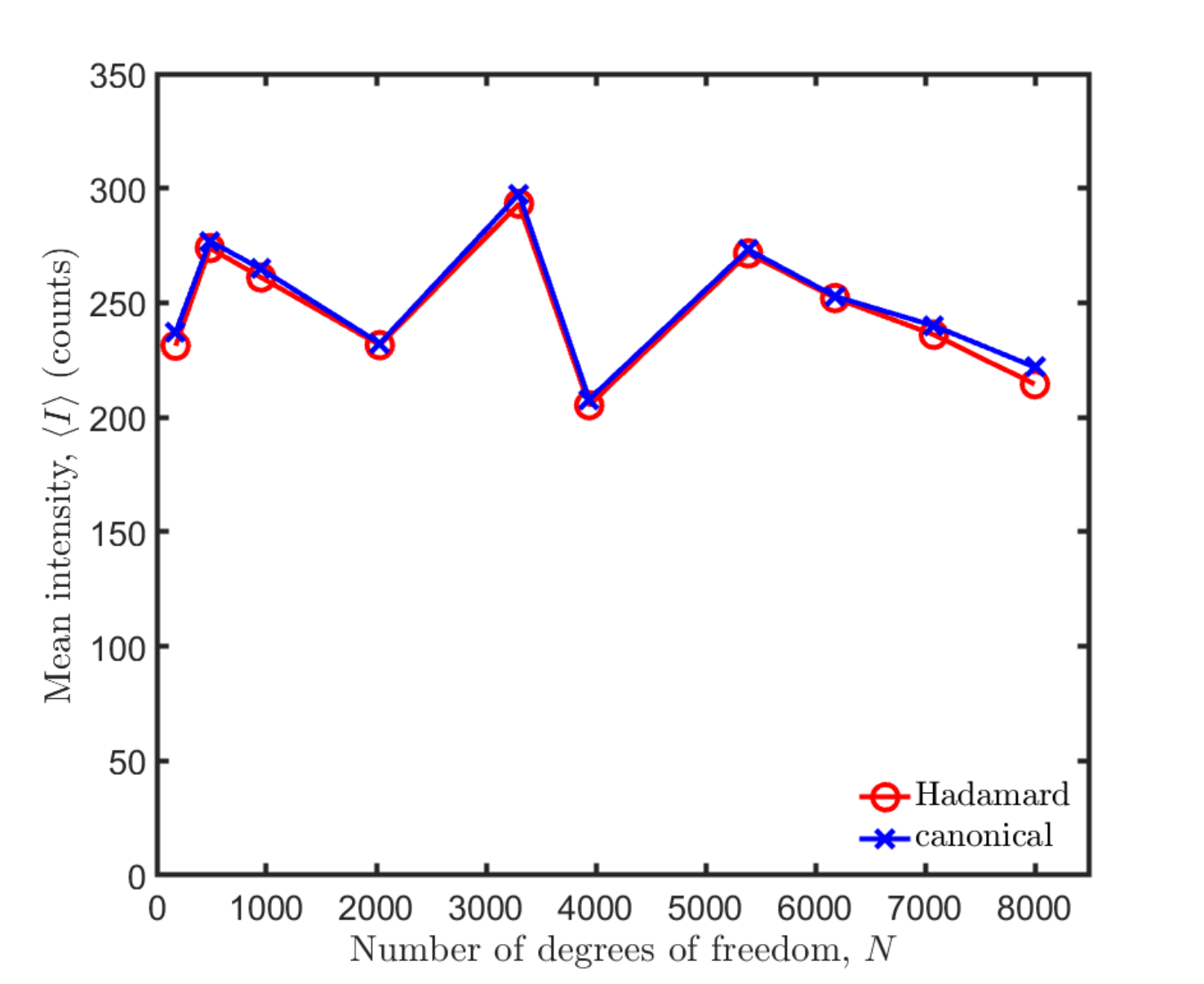}
	\caption{Mean intensity $\langle I \rangle$ (averaged over CCD pixels $m$) recorded on the CCD camera as a function of the number of degrees of freedom $N$ in transmission-matrix measurements using the Hadamard and canonical bases is shown. The red line represents measurements obtained with the Hadamard basis, while the blue line corresponds to the canonical basis. Neutral density (ND) filters were used during the transmission-matrix acquisition process to maintain comparable intensity levels across all measurements.}
	\label{figure8}
\end{figure}

In addition, for all transmission-matrix measurements, we applied a fixed random phase pattern on the SLM with $6 \times 6$ pixel-sized segments, along with the binary phase grating. The displayed fixed random phase pattern ensured uniform coupling of light into all fiber modes, resulting in fully-developed speckle patterns at the fiber's distal end. The modulated beam is reflected off the SLM, passes through the beam splitter, and is Fourier transformed by a lens ($f_1 = 150$ mm). A diaphragm (spatial filter) in the Fourier plane blocks the zeroth-order diffracted light. The filtered light is then relayed through another lens ($f_2 = 150$ mm), Fourier-transformed onto the back aperture of the $\text{MO}_1$, and coupled into the MMF using $\text{MO}_1$ (Olympus Plan $10\times$, NA = 0.25). The fiber is a $6$-cm-long step-index MMF with a core diameter of 200 \textmu m and a numerical aperture NA = 0.22. It supports approximately 15,178 modes per polarization at $\lambda = 561$ nm. To maintain stability, the experimental setup is enclosed in a protective box, and room temperature is controlled to within $\pm 1 \rm\ C^{\circ}$ to mitigate external factors that could affect the accuracy of the transmission-matrix measurements.

To verify the alignment between the SLM input patterns and the MMF core, we observe the reflection from the fiber’s proximal end. A portion of the incident light is back-reflected, and the reflected light pattern within the core at the proximal end is imaged onto  CCD2 with the $\text{MO}_1$ and a lens with a focal length of $f_3 = 300$ mm, enabling the evaluation of the overlap between the displayed SLM patterns and the fiber core. This reflection measurement allows for precise adjustments to optimize coupling efficiency. Simultaneously, the transmitted wavefront propagating through the MMF is imaged at the distal end using the microscope objective ($\text{MO}_2$, Nikon Plan Fluor $10\times$, NA = 0.30) and a lens with a focal length of $f_4 = 300$ mm onto CCD1 (AVT Manta G-040B). By analyzing both the reflected and transmitted intensity patterns, this setup facilitates precise alignment, ensuring efficient coupling and precise control of light propagation.

We measured the reflection at the proximal end of the MMF under different spatial phase modulations applied to the SLM to examine their effect on the intensity patterns reflected. Fig.~\ref{figure7} presents experimental images of the reflected light intensity patterns for different incident wavefronts. Figs.~\ref{figure7}(a)–(c) show the reflections when horizontal grating, vertical grating, and random wavefront, respectively, are applied on the SLM with $6 \times 6$ pixel-sized segments. Similarly, Figs.~\ref{figure7}(d)–(f) display the corresponding reflections for segments with a greater segment size of $12 \times 12$ pixels. The observed reflection patterns confirm the precise alignment of the SLM patterns with the MMF core, ensuring efficient light coupling. To further validate the reliability of our transmission-matrix retrieval process, we analyze the mean output intensity across different wavefront shaping experiments.

Fig.~\ref{figure8} presents the mean output intensity $\langle I \rangle$ over the CCD pixels $m$ as a function of the number of degrees of freedom $N$ for both the Hadamard and canonical bases. During the transmission-matrix acquisition process, we carefully adjust neutral density (ND) filters to ensure that the mean output intensity remains consistent across all measurements. This guarantees that the signal-to-noise level in all transmission-matrix measurements remains comparable.

\subsection*{1. Polarization mixing}

\begin{figure}[H]
	\centering
	\includegraphics[width=9cm]{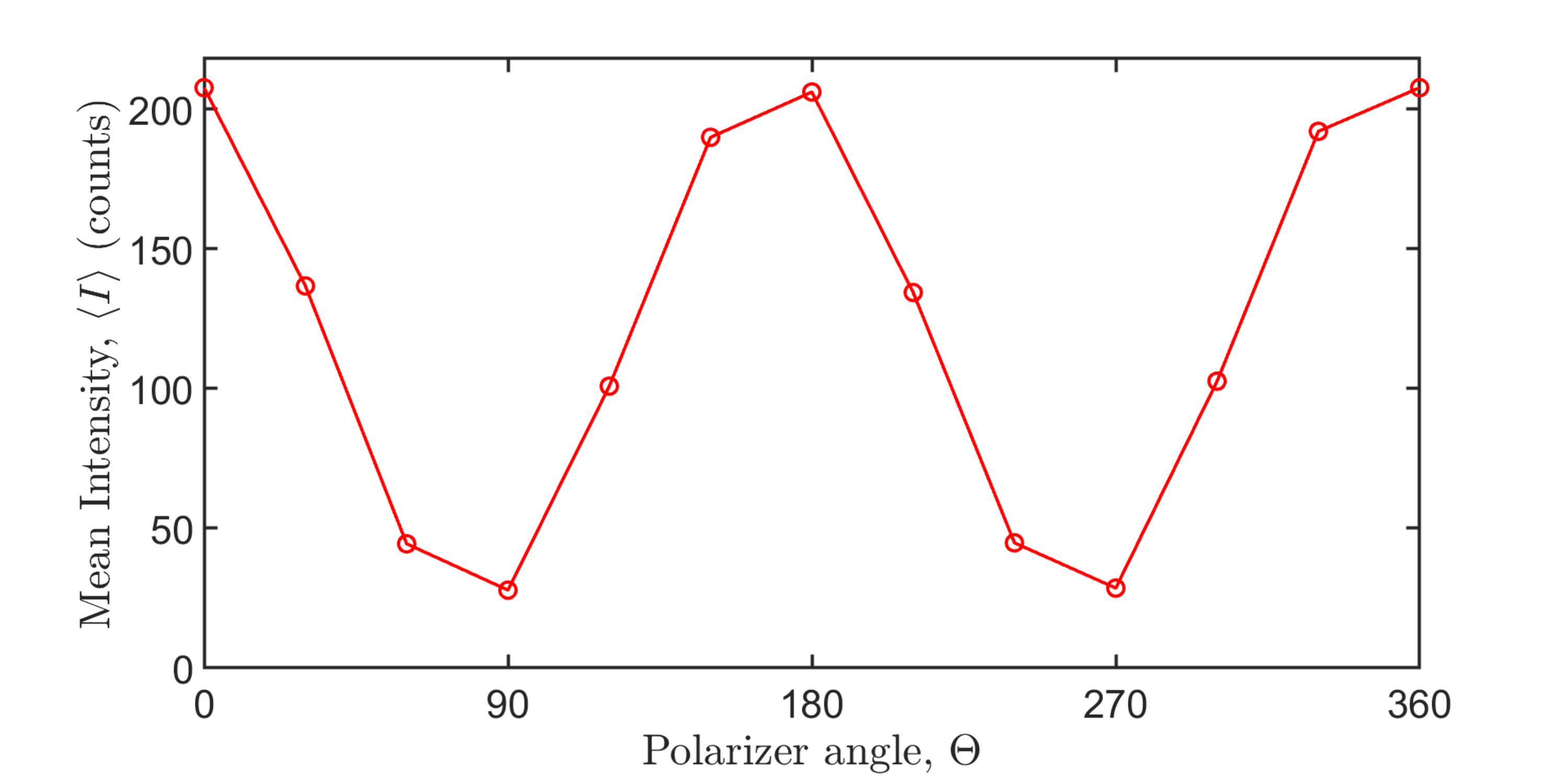}
	\caption{Polarization mixing measurement results are shown. The open red circles show the measured mean intensity $\langle I\rangle$ as a function of the linear polarizer rotation angle $\Theta$. The polarizer is rotated in discrete steps, and the intensity variation indicates that a minimal cross-polarization state is present in the system. The intensity reaches a maximum at angular positions corresponding to multiples of $\pi$ and a minimum at $\Theta = 90^\circ$, confirming the suppression of cross-polarization state.}
	\label{figure9}
\end{figure}

\begin{figure*}[th]
	\centering
	\includegraphics[width=16cm]{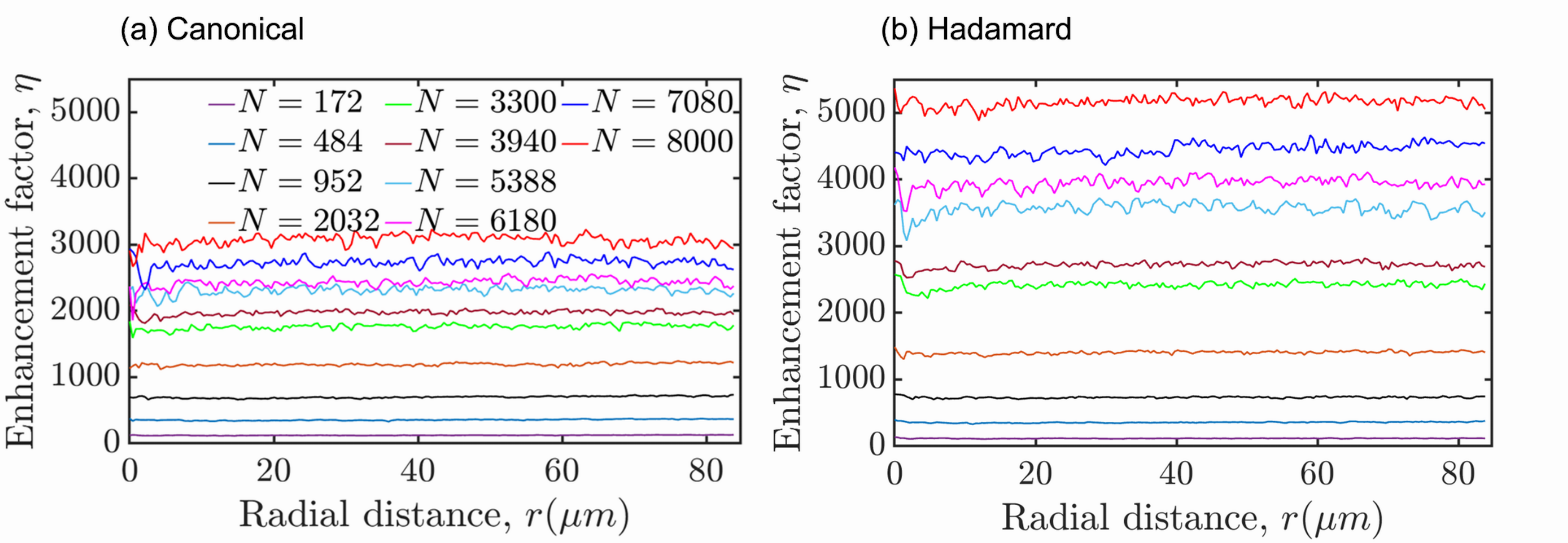}
	\caption{Mean enhancement factor, $\eta$ averaged over azimuthal $\theta$ positions, as a function of the radial distance $r$ at the fiber's distal end for different numbers of degrees of freedom is shown: $N = 172, 484, 952, 2,032, 3,300, 3,940, 5,388, 6,180, 7,080, 8,000$. Enhancement factors obtained from wavefront shaping (a) on the canonical basis, and (b) on the Hadamard basis. The enhancement factors are consistently higher when using the Hadamard basis for transmission-matrix measurement.}
	\label{figure10}
\end{figure*}

In our experimental setup, we place linear polarizers at the input and output of the multimode fiber (MMF) to selectively measure a single polarization of light. While the input beam is horizontally polarized, the polarizer at the output is rotated in increments of $30^\circ$, and the transmitted intensity is recorded on the CCD camera. 

As shown in Fig.~\ref{figure9}, the measured intensity reaches a maximum at angles corresponding to multiples of $\pi$ and drops to a minimum at $\Theta = 90^\circ$, indicating negligible intensity in the vertical cross-polarization state. This confirms that polarization mixing in our fiber is negligible.

\subsection*{2. Experimental enhancement factor}

Fig.~\ref{figure10} shows the enhancement factor $\eta$, averaged over azimuthal position $\theta$, as a function of the radial distance $r$ at the fiber's distal end for wavefront shaping experiments conducted on both the canonical and Hadamard basis. The enhancement factors are measured for various numbers of degrees of freedom, ranging from $N = 172$ to $N = 8,000$.

In Fig.~\ref{figure10}(b), the enhancement factors obtained using the dual-reference algorithm (in the Hadamard basis) are displayed. The dual-reference algorithm yields higher enhancement than the stepwise sequential algorithm (in the canonical basis) due to improved interferometric visibility~\cite{mastiani2021noise}, leading to an optimized signal-to-noise ratio and, consequently, higher enhancement. Comparing the two bases, we find that the measured enhancement factor for wavefront shaping on the Hadamard basis at the largest $N$ is approximately 40\% higher than that obtained on the canonical basis. This difference arises because the signal-to-noise ratio in the Hadamard basis is inherently higher than in the canonical basis, resulting in more accurate transmission-matrix measurements and improved enhancement factor values.

\subsection*{3. Phase-error estimation}

\begin{figure*}[th]
	\centering
	\includegraphics[width=18cm]{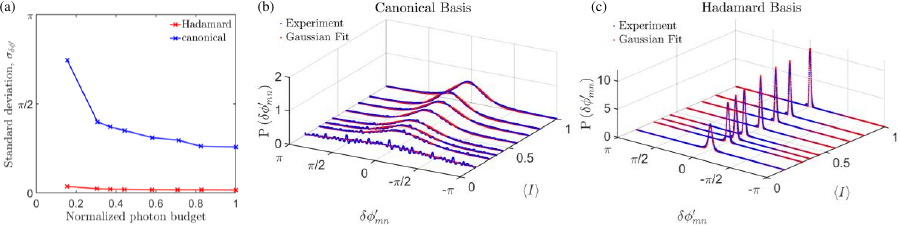} 
	\caption{(a) Standard deviation $\sigma_{\delta\phi^\prime}$ of the phase errors as a function of the normalized photon budget for the phase measurements in canonical and Hadamard basis is shown. The phase measurements in the canonical basis exhibit significantly larger phase errors than in the Hadamard basis, particularly at low photon budget values. The phase-error probability densities for the canonical (b) and Hadamard (c) bases are illustrated. The blue curves represent the experimentally obtained phase-error probability densities, while the red curves correspond to the Gaussian fits used to estimate the standard deviation $\sigma _{\delta\phi^\prime}$. The phase error $\delta \phi_{mn}'$  is computed as the difference between two independently retrieved output phase maps when the same input phase pattern is displayed twice on the spatial light modulator (SLM). The fitted Gaussian function provides the standard deviation $\sigma _{\delta\phi^\prime}$, which quantifies phase fluctuations across different input degrees of freedom $N$.}
	\label{figure11}
\end{figure*}

Phase errors are an inherent aspect of wavefront shaping experiments, arising from variations between the original phase and the measured phase. To quantify this variation, we define the phase error as:

\begin{equation}
\delta \phi_{mn}' = \arg \left({\rm e}^{i(\phi_{mn}^{(2)} - \phi_{mn}^{(1)})} \right)
\end{equation}
where $\phi_{mn}^{(1)}$ and $\phi_{mn}^{(2)}$ represent two independent measurements of the same input phase pattern. To statistically analyze these errors, we generate histograms of $\delta \phi_{mn}'$ and fit them with MATLAB's \texttt{gauss1} function, which models the distribution as:
\begin{equation} 
P(\delta \phi_{mn}') = b \exp\left(-\left(\frac{\delta \phi_{mn}' - \mu}{\sigma_{\delta\phi^\prime}}\right)^2 \right) 
\end{equation}
where $\sigma_{\delta\phi^\prime} $ denotes the standard deviation of the phase-error distribution, and $\mu$ represents its mean. The Gaussian fit enables us to extract $\sigma_{\delta\phi^\prime}$, providing a quantitative measure of phase fluctuations. Next, we numerically generate rescaled Gaussian-distributed random phase values $\phi_{mn}$ with mean $\mu = 0$ and standard deviation $\sigma_{\delta\phi^\prime}/\sqrt{2}$. Using these rescaled random phase errors, we compute the phase-error coefficient:
\begin{equation}
\Phi_m \equiv \left\langle \cos(\phi_{mn}) \right\rangle_n^2
\end{equation}
where the averaging is performed over all inputs $n$. This coefficient quantifies the impact of input phase errors on the performance of wavefront shaping. As shown in Fig.~\ref{figure11}, the phase-error standard deviations exhibit significant differences between the canonical and Hadamard bases. The probability density of phase errors in the Hadamard basis is sharper and more concentrated, indicating reduced phase errors compared to the canonical basis. Fig.~\ref{figure11}(a) presents the standard deviation $\sigma_{\delta\phi^\prime} $ of the phase errors as a function of the normalized photon budget. The canonical basis consistently exhibits higher phase errors across all photon budgets, particularly in low-light conditions where the measurement noise dominates. In contrast, the Hadamard basis results in significantly lower phase errors, demonstrating its robustness in maintaining accurate phase retrieval even under reduced photon budgets. These results highlight the influence of basis selection in reducing phase errors, with potential implications for optimizing wavefront shaping performance in photon-limited scenarios.

\section*{Appendix B: Theory and Numerical Simulations}

\subsection*{1. Derivation of the enhancement factor $\eta$}

Here, we present the derivation of the enhancement factor $\eta$, following the approach in Ref.~\cite{vanBeijnum2008thesis} for focusing light through a complex medium.  
The field at the output channel $m$ from $N$ incident channels is given by
\begin{align}
    E_m = \sum_{n = 1}^N t_{mn}E_n,
\end{align}
where $t_{mn}$ is the transmission-matrix element and $E_n$ is the incident field in channel $n$.  
In general, both the amplitude and phase of $E_n$ can be controlled. Our goal is to determine the optimal amplitude and phase configuration to maximize the intensity in a single target output channel $m$.

The intensity in channel $m$ is
\begin{align}
    I_m &= \left| \sum_{n = 1}^N t_{mn}A_n e^{\phi_n} \right|^2 \\
        &= \sum_{n = 1}^N |t_{mn}|^2 A_n^2
        + \sum_{n = 1}^N \sum_{n' \neq n}^N t_{mn}A_n e^{\phi_n} \, t_{mn'}^* A_{n'} e^{-\phi_{n'}},
    \label{eq:int_mean}
\end{align}
where $A_n$ and $\phi_n$ are the amplitude and phase of the $n$th input channel.

In practice, $I_m$ cannot be evaluated meaningfully for a single realization of $t_{mn}$ or $E_n$.  
For disordered scattering media, averaging can be performed over different disorder realizations, i.e., over $t_{mn}$.  
However, in a single multimode fiber, $t_{mn}$ is fixed and does not vary statistically across fibers, so averaging over fiber realizations is not meaningful.  

Instead, we perform an average over many independent random input fields $E_n^{\mathrm{(rand)}}$, with each input field having complex elements that are independent and identically distributed according to a Gaussian distribution. For such independent random inputs, the cross term in Eq.~\eqref{eq:int_mean} vanishes:
\begin{align}
    \sum_{n = 1}^N \sum_{n' \neq n}^N t_{mn}A_n e^{\phi_n} \, t_{mn'}^* A_{n'} e^{-\phi_{n'}} = 0,
\end{align}
yielding
\begin{align}
    \langle I_m \rangle = N \, \langle |t_{mn}|^2 \rangle_n \, \langle A_n^2 \rangle_n.
\end{align}

The maximum possible intensity is obtained when the phases $\phi_n$ are chosen for perfect constructive interference in channel $m$:
\begin{align}
    I_{m,\mathrm{max}} &= \left| \sum_{n = 1}^N t_{mn}A_n e^{\phi_n} \right|^2 \\
        &= \sum_{n = 1}^N |t_{mn}|^2 A_n^2
        + \sum_{n = 1}^N \sum_{n' \neq n}^N |t_{mn}|A_n \, |t_{mn'}|A_{n'},
    \label{eq:int_max}
\end{align}
where the second term is maximized when all contributions are in phase.

If $A_n$ is statistically independent of $|t_{mn}|$, $|t_{mn'}|$, and $A_{n'}$, then averaging Eq.~\eqref{eq:int_max} gives
\begin{align}
    \langle I_{m,\mathrm{max}} \rangle = N \langle |t_{mn}|^2 \rangle_n \langle A_n^2 \rangle_n
    + N(N-1) \langle |t_{mn}| \rangle_n^2 \langle A_n \rangle_n^2.
\end{align}

The average enhancement factor $\eta$ is the ratio of the average optimized intensity to the average random-speckle intensity:
\begin{align}
    \eta = 1 + (N-1) \frac{\langle |t_{mn}| \rangle_n^2}{\langle |t_{mn}|^2 \rangle_n}
    \frac{\langle A_n \rangle_n^2}{\langle A_n^2 \rangle_n}.
\end{align}
For phase-only modulation with a spatially uniform input amplitude ($A_n = \text{const}$), the ratio $\langle A_n \rangle_n^2 / \langle A_n^2 \rangle_n = 1$, giving
\begin{align}
    \eta = \left[ \frac{\langle |t_{mn}| \rangle_n^2}{\langle |t_{mn}|^2 \rangle_n} \right] (N-1) + 1,
\end{align}
where
\begin{align}
    R_m = \frac{\langle |t_{mn}| \rangle_n^2}{\langle |t_{mn}|^2 \rangle_n}
\end{align}
is the participation ratio.

\subsection*{2. Step-index fiber}

We consider step-index (SI) multimode fibers with radius $a$, with a core of refractive index $n_1$ and a cladding of refractive index $n_2$, described by the radial refractive index profile:
\begin{align}
\begin{array}{ll}
\displaystyle n(r) = \begin{cases} 
 n_1, & \text{ $0\le r\le a$} \\  
n_2, & \text{$r > a$}.
 \end{cases} 
\end{array} 
\end{align}
To numerically define the transmission matrix, we employ mode decomposition, meaning that the transmission matrix consists of linearly polarized (LP) modes $\psi_{l,p}(r,\theta)$ with radial mode number $l$ and azimuthal mode number $p$ arranged in its columns~\cite{Lee2023GitHub}. The characteristic equations for LP modes involve the radial functions within the core $(h_a)$ and in the cladding $(q_a)$, given by,
\begin{align}
    h_a & = a\sqrt{(n_1k_0)^2-\beta^2},\\
    q_a & =  a\sqrt{\beta^2-(n_2k_0)^2},
\end{align}
where $a$ denotes the core radius, $n_1$ the core refractive index, $n_2$ the cladding refractive index, $k_0 = 2\pi/\lambda$ the vacuum wave number, and $\beta$ propagation constants. The LP mode characteristic equation combines these radial functions with Bessel functions for the core $J_l$ and modified Bessel functions for the cladding $K_l$, structured as follows;
\begin{equation}
    \\h _ a \left( \frac{J _ {l+1}(h _ a )}{J_l(h _ a )} \right) = q _ a  \left( \frac{K _ {l+1}(q _ a )}{K_l(q _ a )} \right).
\end{equation}

We solve the transcendental equation numerically and obtain the propagation constants $\beta$ for weakly guided LP modes in the multimode fiber. Furthermore, the transmission matrix can be constructed through a mode matrix $\psi$ containing each  LP mode with a given $l$ and $p$ $\psi_{l,p}(r,\theta)$ within its columns. Consequently, we can express the real-space transmission matrix that maps the input field on the proximal end position $s$ to the output field on the distal end position $m$ as follows;
\begin{equation}
    t_{m s} = \psi_{m l,p}e^{i\beta_{l,p} L}\psi_{sl,p}^\dagger,
\end{equation}
where $\beta_{l,p}$ is the propagation constant for the mode with radial mode number $l$ and azimuthal mode number $p$, and $L$ the fiber length. 

In our numerical simulations, we compute the transmission matrices for SI fibers with a core diameter of 25 \textmu m and a length of 0.1 m at a wavelength of $\lambda = 561$ nm. The numerical aperture is ${\rm NA} = 0.22$. The $V$ number for step-index fiber is given by
\begin{equation}
    V_{\rm SI} = k_0a{\rm NA},
\end{equation}
where the approximate number of guided modes per polarization is
\begin{equation}
    M_{\rm SI} = \frac{V_{\rm SI}^2}{4}.
\end{equation}

We restrict the numerical simulations to a single linear polarization. This is because we are using linearly polarized input in the experiment, and there is no significant polarization mixing as illustrated in Fig.~\ref{figure9}.

\subsection*{3. Graded-index fiber}

We consider graded-index (GRIN) multimode fibers that support multiple modes, featuring a core of radius $a$ with a parabolic refractive index profile and a cladding of constant refractive index $n_2$. The radial refractive index profile is given by:
\begin{align}
\begin{array}{ll}
\displaystyle n(r) = \begin{cases} 
 n_1\sqrt{1-2\Delta n\frac{r^2}{a^2}}, & \text{ $0\le r\le a$} \\  
n_2, & \text{$r \ge a$}.
 \end{cases} 
\end{array} 
\end{align}
\begin{align}
\Delta n = \frac{n_1^2-n_2^2}{2n_1^2}
\end{align}
 is the relative index difference, For $\Delta n \ll 1$, this reduces to
 \begin{align}
 \Delta n \approx \frac{n_1-n_2}{n_1}. 
 \end{align}
\begin{figure*}[th]
	\centering
	\includegraphics[width=14cm]{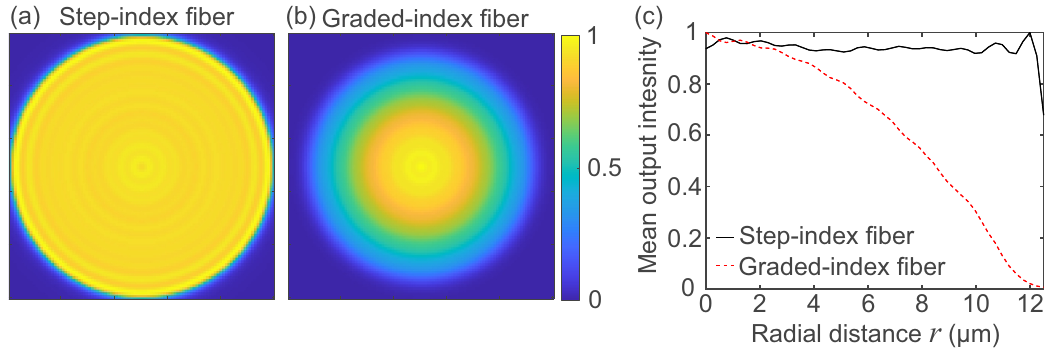}
	\caption{Mean output intensity profiles $\langle |t_{ml,p}|^2 \rangle_{l, p}$ within the core at the distal end of (a) a step-index (SI) fiber and (b) a graded-index (GRIN) fiber, both with a core diameter of 25 \textmu m, are shown. (c) Azimuthally averaged mean intensity profiles are also shown, with the black solid line representing $\langle |t_{ml,p}|^2 \rangle_{l, p}$ for the step-index fiber and the red dashed line representing $\langle |t_{ml,p}|^2 \rangle_{l, p}$ for the graded-index fiber.}
	\label{figure12}
\end{figure*}
 Solving the scalar Helmholtz equation in a straight waveguide leads to the identification of scalar mode profiles $\psi_{l,p}(r,\theta)$, characterized by a radial mode number $l$ and an azimuthal mode number $p$. In GRIN fibers, these mode profiles take the form of Laguerre-Gaussian (LG) functions, with the same mode indices $l$ and $p$~\cite{boonzajer2018robustness}. Then, we express the transmission matrix of the GRIN fiber using the LG modes and propagation constants
\begin{align}
\beta_{l,p} = \sqrt{k_0^2 n_1^2 - 2\alpha (|p| + 2l + 1)},
\end{align}
where $k_0$ is the vacuum wavenumber, $n_1$ is the axial core refractive index, $l$ and $p$ denote the radial and azimuthal mode numbers, respectively, and
\begin{equation}
\alpha = \frac{k_0 n_1 \sqrt{2\Delta n}}{a}.
\end{equation}
The $V$ number for a GRIN fiber with a parabolic refractive index profile is given by
\begin{equation}
    V_{\rm GRIN} = k_0 a n_1 \sqrt{2 \Delta n},
\end{equation}
where the approximate number of guided modes per polarization is 
\begin{equation}
    M_{\rm GRIN} = \frac{V_{\rm GRIN}^2}{8}.
\end{equation}

In our numerical simulations, we compute the transmission matrix for a GRIN fiber with a core diameter of $25\,\text{\textmu m}$ and a length of $0.1\,\text{m}$ at a wavelength of $\lambda = 561\,\text{nm}$. The refractive index at the core center is $n_1 = 1.480$, and the cladding refractive index is $n_2 = 1.45422$.

\subsection*{4. Mean output intensity}

The mean output intensity profiles for both the SI and GRIN fibers are calculated by averaging the transmission-matrix elements over all input modes $l, p$, given by $\langle |t_{ml,p}|^2 \rangle_{l, p}$. The SI fiber supports $M_{\mathrm{SI}} = 244$ modes per polarization, and the GRIN fiber supports $M_{\mathrm{GRIN}} = 190$ modes per polarization. The mean output intensity profiles are shown in Fig.~\ref{figure12}. It is important to note that the mean output intensity is not a direct measure of mode density. Instead, mode density is quantified by the participation ratio $\langle R_{m l,p} \rangle_{l,p} = \langle |t_{m l,p}| \rangle_{l,p}^2 / \langle |t_{m l,p}|^2 \rangle_{l,p}$, which represents the effective number of modes contributing to the field at a given position within the core. We provide a detailed description of the participation ratio in the next subsection.

\subsection*{5. Participation ratio}

\begin{figure*}[th]
	\centering
	\includegraphics[width=14cm]{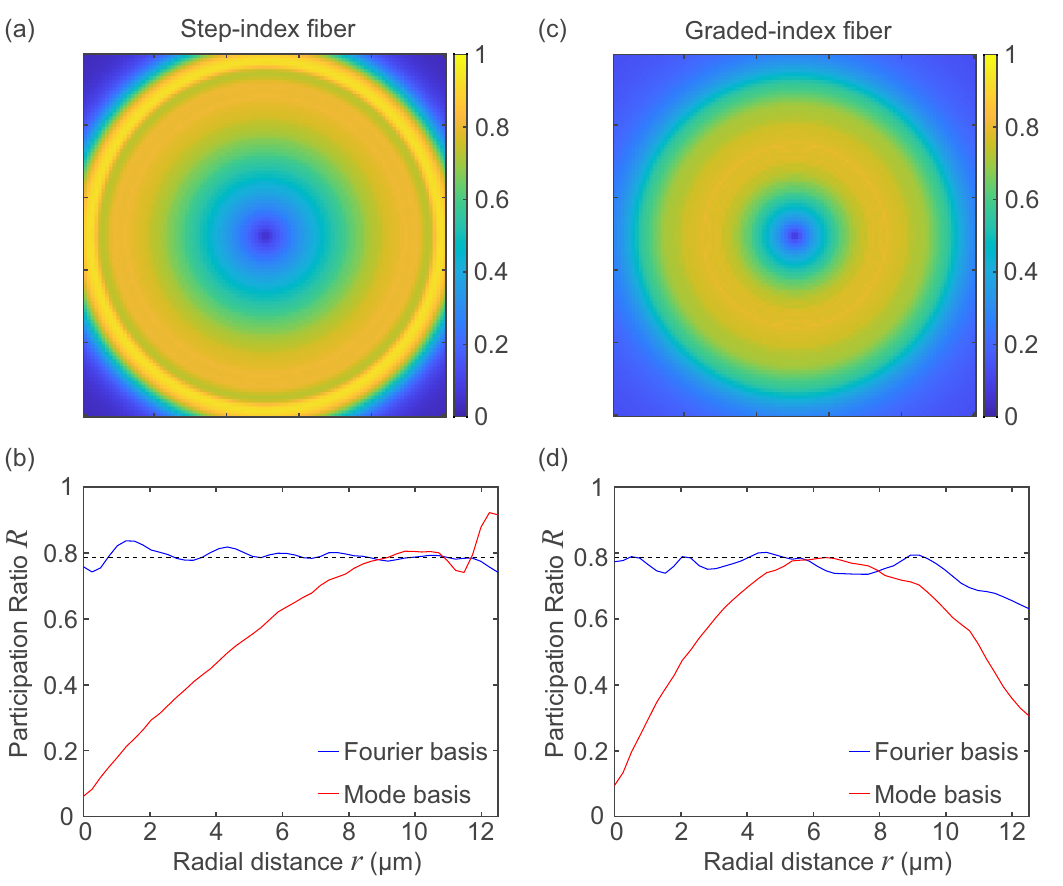}
	\caption{Participation ratio for mode basis input (mode density) $\langle R_{m l,p}\rangle_{l,p}$ and the radial distance $r$ dependence of $\langle R_{m l,p}\rangle_{l,p}$ and $\langle R_{mn}\rangle_{n}$ within the core at the distal end of (a,b) a step-index fiber and (c,d) a graded-index fiber, both with a core diameter of 25 \textmu m, are shown. When input phase-only wavefront modulation is applied in the Fourier basis ($\langle R_{mn}\rangle_{n}$), both $R$ values tend to converge to $R = \pi/4$, except at the edge of the core in the graded-index fiber, where $R$ drops to approximately 0.6. In contrast, phase-only modulation in the mode basis ($\langle R_{m l,p}\rangle_{l,p}$) yields $R < 0.2$—at the center of the core for both the step-index fiber and the graded-index fiber. Moreover, the participation ratio $R$ decreases to approximately $R \approx 0.3$ at the edge of the core on the distal end of the graded-index fiber. The blue and red solid lines represent the participation ratio $R$ in the Fourier basis and mode basis, respectively, at the input. Black dashed lines indicate the value $\pi/4$.}
	\label{figure13}
\end{figure*}

The participation ratio $R$ is a key metric that captures how many independent input degrees of freedom effectively contribute to the optical field at a given output position. Unlike mere intensity measurements, which offer no insight into the underlying mode structure, the participation ratio reveals the true modal diversity involved in light focusing—providing a bridge between wavefront shaping and the spatial structure of fiber modes. When the participation ratio is computed for input modulation in the mode basis, it directly corresponds to the mode density at the target output position. To determine the participation ratio for input wavefronts in the Fourier and mode bases, as well as the output in real space at the distal end core surface, we compute and decompose the fiber transmission matrix into an appropriate basis.

The transmission matrix that quantifies the mapping between the input fields in mode basis to output fields on the distal end is represented by
\begin{equation}
    t_{m l,p} = \psi_{m l,p}e^{i\beta_{l,p} L}.
\end{equation}
Next, we compute the participation ratio $R_m$ for a certain fiber output position $m$ at the distal end when input modulation is in the fiber mode basis as
\begin{equation}\label{corrfunc}
    \langle R_{m l,p}\rangle_{l,p} = \frac{\langle|t_{m l,p}|\rangle_{l,p}^2}{\langle|t_{m l,p}|^2\rangle_{l,p}}.
\end{equation} 

To compute the participation ratio $R_m$, when phase-only modulation is applied in the Fourier plane, we perform a two-dimensional Fourier decomposition on each row $s$ of the real-space transmission matrix, and obtain
\begin{equation}
    t_{mn} = t_{ms}D_{n s}^\dagger,
\end{equation}
where $D_{n s}$ represents a two-dimensional discrete Fourier decomposition matrix.

Finally, we compute the participation ratio $R_m$ for a certain fiber output at the distal end $m$ when input modulation is in the Fourier basis as
\begin{equation}\label{corrfunc}
    \langle R_{m n}\rangle_n = \frac{\langle|t_{mn}|\rangle_{n}^2}{\langle|t_{mn}|^2\rangle_{n}}.
\end{equation}
The participation ratios $R$ as a function of radial distance within the fiber core at the distal end are shown in Fig.~\ref{figure13} for both step-index and graded-index fibers, under input modulation in the Fourier and mode bases.

\subsection*{6. Digital optical phase conjugation and the enhancement factor $\eta$}

\begin{figure*}[th]
	\centering
	\includegraphics[width=14cm]{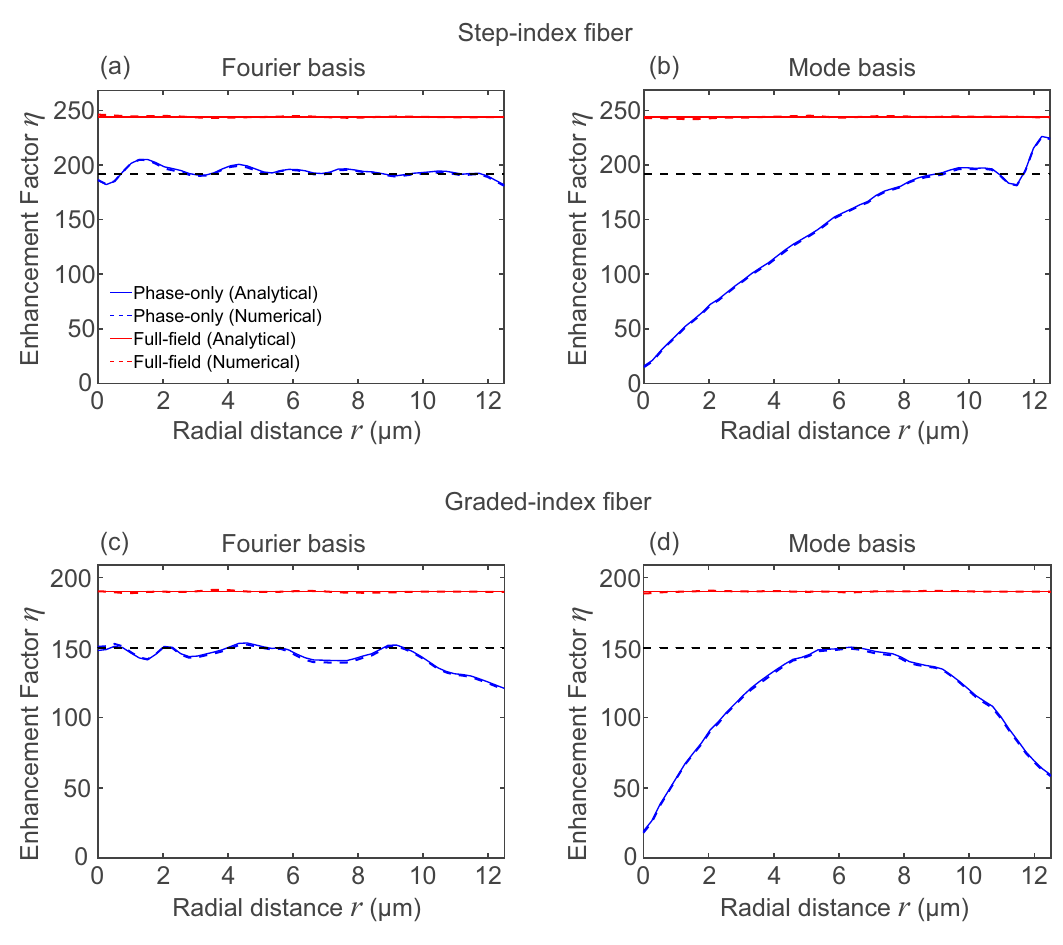}
	\caption{The enhancement factor $\eta$ as a function of radial distance $r$ within the core at the distal end of (a,b) a step-index fiber and (c,d) a graded-index fiber, both with a core diameter of 25 \textmu m, is shown. Phase-only input modulation in the Fourier basis (a,c) results in a nearly uniform enhancement across both fibers’ cores, with $\eta$ approaching $\eta = (\pi/4)(N-1) + 1$, except for a noticeable drop at the edge of the core in the graded-index fiber. In contrast, phase-only modulation in the mode basis (b,d) leads to a strong radial dependence of $\eta$ across both fibers’ cores, with $\eta$ minimized at the core center for both fibers. Morevover, the enhancement factor $\eta$ decreases at the edge of the core on the distal end of the graded-index fiber. For comparison, full-field (amplitude and phase) modulation yields a uniform enhancement of $\eta = N$ in all cases, where $N = 244$ and $N = 190$ for step-index and graded-index fiber, respectively. Blue solid and dashed lines represent the analytical and numerical values of $\eta$ for phase-only input modulation, respectively; red solid and dashed lines correspond to the analytical and numerical values of $\eta$ for full-field input modulation. The black dashed line indicates $\eta = (\pi/4)(N-1) + 1$.}
	\label{figure14}
\end{figure*}

In this section, we present our numerical and analytical results for the enhancement factor when focusing light through step-index (SI) and graded-index (GRIN) multimode fibers (MMFs).

For the numerical calculations, we use transmission matrices obtained from simulations of both SI and GRIN fibers. Focusing simulations are performed by applying wavefront shaping at the input, using either the Fourier basis or the mode basis. We compute the numerical enhancement factor as a function of the radial distance $r$ within the fiber core at the distal end, for both fiber types. Each simulation is performed for two input modulation schemes—full-field (amplitude and phase) and phase-only—implemented in both the Fourier and mode bases.

To generate the input wavefronts, we employ digital optical phase conjugation (DOPC), which enables the formation of a diffraction-limited focus at a chosen location $m$ within the fiber core at the output. This is achieved by selecting the corresponding row $m$ of the transmission matrix, which contains the output field response for all inputs at the desired focus position $m$.

In the transmission-matrix representation, each column corresponds to the output field distribution produced by a single input mode or Fourier component, while each row represents the input field required to produce a certain field value at a specific output position. Thus, to focus light at the output position $m$ at the distal end of the fiber, we use the complex conjugate of the corresponding row as the input field,
\begin{equation}
    \mathbf{E}^{\text{(in)}}_n = \sum_{m = 1}^{M}t^*_{nm} \mathbf{E}_m^{\text{(out)}},
\end{equation}
where $t^{*}$ represents the complex conjugate of the transmission matrix, and $\mathbf{E}_m^{\text{(out)}}$ corresponds to the desired field at the output. In case one wishes to focus light onto a single diffraction-limited spot—corresponding to the $m{\rm th}$ output channel—the desired output field $\mathbf{E}_m^{\text{(out)}}$ is represented by a column vector of size $M \times 1$ containing a one at the $m{\rm th}$ position and zeros elsewhere:
\begin{align}
    \mathbf{E}^{\text{(out)}}_m &= \begin{bmatrix}
           \vdots \\
           0 \\
           1 \\
           0  \\
           \vdots
         \end{bmatrix}.
  \end{align}
This vector specifies that all energy is to be concentrated at the $m{\rm th}$ output channel, with no intensity at the others. We use $\mathbf{E}^{\text{(in)}}_n$ as the full-field (amplitude and phase) input, and ${\arg}[\mathbf{E}^{\text{(in)}}_n]/\sqrt{N}$ as the normalized phase-only input field, both of which are multiplied by the transmission matrix from the right-hand side. We then compute the numerical enhancement factor as
\begin{equation}
\eta_{m} = \frac{|t_{mn}\mathbf{E}^{\text{(in)}}_n|^2}{\langle |t_{mn}|^2 \rangle_n},
\end{equation}
where $\eta_m$ quantifies the enhancement at output channel $m$ relative to the mean transmission intensity over all input channels $n$. We use this expression to compute the numerical enhancement factors shown in Fig.~\ref{figure14}.

The general form of the theoretical enhancement factor for focusing through a complex medium with phase-only input modulation—assuming no phase errors—is given by
\begin{equation}
    \eta_{m} = R_m(N-1)+1,
\end{equation}
where $m$ denotes the focusing position index at the output, such as the radial distance on the fiber core at the distal end. We use this expression to compute the analytical enhancement factors shown in Fig.~\ref{figure14}, where the participation ratio is defined as $R \equiv \langle R_{mn} \rangle_n$ for input modulation in the Fourier basis, and as $R \equiv \langle R_{m l,p} \rangle_{l,p}$ for input modulation in the mode basis. It is important to emphasize that this expression is general. The participation ratio $R_m$ must be calculated using the transmission matrix of the complex medium. For example, for a scattering medium, $R = \pi/4$, whereas for a step-index or graded-index multimode fiber, $R_m$ depends on the specific basis used at the modulation input plane.

For full-field modulation, the enhancement factor reaches the theoretical maximum $\eta = N$ in both Fourier and mode bases, and for both SI and GRIN fibers. Under phase-only modulation in the Fourier basis, we observe a nearly uniform enhancement across the entire core, with $\eta$ approaching the analytical value of $(\pi/4)(N-1) + 1$, except for a noticeable drop near the edge of the core in the GRIN fiber.

In contrast, phase-only modulation in the mode basis results in a strong radial dependence of the enhancement factor. For both the SI and GRIN fibers, $\eta$ is minimized at the center of the core. Furthermore, for the GRIN fiber, the enhancement factor drops again near the core edge.

This radial dependence can be understood through the participation ratio $R$, which quantifies the number of input degrees of freedom contributing to the field at the target output position $m$, as shown in Fig.~\ref{figure13}. In the Fourier basis, each input component excites a broad superposition of fiber modes, resulting in a uniform spatial contribution across the core. However, in the mode basis, each input corresponds directly to a single fiber mode, and the spatial distribution of these modes varies significantly.

In both SI and GRIN fibers, most modes exhibit ring-shaped profiles that avoid the core center, resulting in a low mode density—and consequently a low participation ratio—at that location. This accounts for the reduced enhancement at the core center. In GRIN fibers, the refractive index gradient supports modes that are concentrated between the core center and the core edge. As a result, both the mode density and participation ratio are lower not only at the center but also near the edge of the core, leading to reduced enhancement in these regions.
	
\bibliography{fiber_focusing}

\begin{thebibliography}{62}%
\makeatletter
\providecommand \@ifxundefined [1]{%
 \@ifx{#1\undefined}
}%
\providecommand \@ifnum [1]{%
 \ifnum #1\expandafter \@firstoftwo
 \else \expandafter \@secondoftwo
 \fi
}%
\providecommand \@ifx [1]{%
 \ifx #1\expandafter \@firstoftwo
 \else \expandafter \@secondoftwo
 \fi
}%
\providecommand \natexlab [1]{#1}%
\providecommand \enquote  [1]{``#1''}%
\providecommand \bibnamefont  [1]{#1}%
\providecommand \bibfnamefont [1]{#1}%
\providecommand \citenamefont [1]{#1}%
\providecommand \href@noop [0]{\@secondoftwo}%
\providecommand \href [0]{\begingroup \@sanitize@url \@href}%
\providecommand \@href[1]{\@@startlink{#1}\@@href}%
\providecommand \@@href[1]{\endgroup#1\@@endlink}%
\providecommand \@sanitize@url [0]{\catcode `\\12\catcode `\$12\catcode `\&12\catcode `\#12\catcode `\^12\catcode `\_12\catcode `\%12\relax}%
\providecommand \@@startlink[1]{}%
\providecommand \@@endlink[0]{}%
\providecommand \url  [0]{\begingroup\@sanitize@url \@url }%
\providecommand \@url [1]{\endgroup\@href {#1}{\urlprefix }}%
\providecommand \urlprefix  [0]{URL }%
\providecommand \Eprint [0]{\href }%
\providecommand \doibase [0]{https://doi.org/}%
\providecommand \selectlanguage [0]{\@gobble}%
\providecommand \bibinfo  [0]{\@secondoftwo}%
\providecommand \bibfield  [0]{\@secondoftwo}%
\providecommand \translation [1]{[#1]}%
\providecommand \BibitemOpen [0]{}%
\providecommand \bibitemStop [0]{}%
\providecommand \bibitemNoStop [0]{.\EOS\space}%
\providecommand \EOS [0]{\spacefactor3000\relax}%
\providecommand \BibitemShut  [1]{\csname bibitem#1\endcsname}%
\let\auto@bib@innerbib\@empty
\bibitem [{\citenamefont {Čižmár}\ and\ \citenamefont {Dholakia}(2012)}]{2012_Cizmar_NC}%
  \BibitemOpen
  \bibfield  {author} {\bibinfo {author} {\bibfnamefont {T.}~\bibnamefont {Čižmár}}\ and\ \bibinfo {author} {\bibfnamefont {K.}~\bibnamefont {Dholakia}},\ }\bibfield  {title} {\bibinfo {title} {Exploiting multimode waveguides for pure fibre-based imaging},\ }\href@noop {} {\bibfield  {journal} {\bibinfo  {journal} {Nat. Commun.}\ }\textbf {\bibinfo {volume} {3}},\ \bibinfo {pages} {1027} (\bibinfo {year} {2012})}\BibitemShut {NoStop}%
\bibitem [{\citenamefont {Choi}\ \emph {et~al.}(2012)\citenamefont {Choi}, \citenamefont {Yoon}, \citenamefont {Kim}, \citenamefont {Yang}, \citenamefont {Fang-Yen}, \citenamefont {Dasari}, \citenamefont {Lee},\ and\ \citenamefont {Choi}}]{2012_Choi_PRL}%
  \BibitemOpen
  \bibfield  {author} {\bibinfo {author} {\bibfnamefont {Y.}~\bibnamefont {Choi}}, \bibinfo {author} {\bibfnamefont {C.}~\bibnamefont {Yoon}}, \bibinfo {author} {\bibfnamefont {M.}~\bibnamefont {Kim}}, \bibinfo {author} {\bibfnamefont {T.~D.}\ \bibnamefont {Yang}}, \bibinfo {author} {\bibfnamefont {C.}~\bibnamefont {Fang-Yen}}, \bibinfo {author} {\bibfnamefont {R.~R.}\ \bibnamefont {Dasari}}, \bibinfo {author} {\bibfnamefont {K.~J.}\ \bibnamefont {Lee}},\ and\ \bibinfo {author} {\bibfnamefont {W.}~\bibnamefont {Choi}},\ }\bibfield  {title} {\bibinfo {title} {Scanner-free and wide-field endoscopic imaging by using a single multimode optical fiber},\ }\href@noop {} {\bibfield  {journal} {\bibinfo  {journal} {Phys. Rev. Lett.}\ }\textbf {\bibinfo {volume} {109}},\ \bibinfo {pages} {203901} (\bibinfo {year} {2012})}\BibitemShut {NoStop}%
\bibitem [{\citenamefont {Papadopoulos}\ \emph {et~al.}(2012)\citenamefont {Papadopoulos}, \citenamefont {Farahi}, \citenamefont {Moser},\ and\ \citenamefont {Psaltis}}]{2012_Papadopoulos_OE}%
  \BibitemOpen
  \bibfield  {author} {\bibinfo {author} {\bibfnamefont {I.~N.}\ \bibnamefont {Papadopoulos}}, \bibinfo {author} {\bibfnamefont {S.}~\bibnamefont {Farahi}}, \bibinfo {author} {\bibfnamefont {C.}~\bibnamefont {Moser}},\ and\ \bibinfo {author} {\bibfnamefont {D.}~\bibnamefont {Psaltis}},\ }\bibfield  {title} {\bibinfo {title} {Focusing and scanning light through a multimode optical fiber using digital phase conjugation},\ }\href@noop {} {\bibfield  {journal} {\bibinfo  {journal} {Opt. Express}\ }\textbf {\bibinfo {volume} {20}},\ \bibinfo {pages} {10583} (\bibinfo {year} {2012})}\BibitemShut {NoStop}%
\bibitem [{\citenamefont {Papadopoulos}\ \emph {et~al.}(2013)\citenamefont {Papadopoulos}, \citenamefont {Farahi}, \citenamefont {Moser},\ and\ \citenamefont {Psaltis}}]{papadopoulos2013high}%
  \BibitemOpen
  \bibfield  {author} {\bibinfo {author} {\bibfnamefont {I.~N.}\ \bibnamefont {Papadopoulos}}, \bibinfo {author} {\bibfnamefont {S.}~\bibnamefont {Farahi}}, \bibinfo {author} {\bibfnamefont {C.}~\bibnamefont {Moser}},\ and\ \bibinfo {author} {\bibfnamefont {D.}~\bibnamefont {Psaltis}},\ }\bibfield  {title} {\bibinfo {title} {High-resolution, lensless endoscope based on digital scanning through a multimode optical fiber},\ }\href@noop {} {\bibfield  {journal} {\bibinfo  {journal} {Biomed. Opt. Express}\ }\textbf {\bibinfo {volume} {4}},\ \bibinfo {pages} {260} (\bibinfo {year} {2013})}\BibitemShut {NoStop}%
\bibitem [{\citenamefont {Loterie}\ \emph {et~al.}(2015)\citenamefont {Loterie}, \citenamefont {Farahi}, \citenamefont {Papadopoulos}, \citenamefont {Goy}, \citenamefont {Psaltis},\ and\ \citenamefont {Moser}}]{loterie2015digital}%
  \BibitemOpen
  \bibfield  {author} {\bibinfo {author} {\bibfnamefont {D.}~\bibnamefont {Loterie}}, \bibinfo {author} {\bibfnamefont {S.}~\bibnamefont {Farahi}}, \bibinfo {author} {\bibfnamefont {I.}~\bibnamefont {Papadopoulos}}, \bibinfo {author} {\bibfnamefont {A.}~\bibnamefont {Goy}}, \bibinfo {author} {\bibfnamefont {D.}~\bibnamefont {Psaltis}},\ and\ \bibinfo {author} {\bibfnamefont {C.}~\bibnamefont {Moser}},\ }\bibfield  {title} {\bibinfo {title} {Digital confocal microscopy through a multimode fiber},\ }\href@noop {} {\bibfield  {journal} {\bibinfo  {journal} {Opt. Express}\ }\textbf {\bibinfo {volume} {23}},\ \bibinfo {pages} {23845} (\bibinfo {year} {2015})}\BibitemShut {NoStop}%
\bibitem [{\citenamefont {Amitonova}\ and\ \citenamefont {De~Boer}(2018)}]{amitonova2018compressive}%
  \BibitemOpen
  \bibfield  {author} {\bibinfo {author} {\bibfnamefont {L.~V.}\ \bibnamefont {Amitonova}}\ and\ \bibinfo {author} {\bibfnamefont {J.~F.}\ \bibnamefont {De~Boer}},\ }\bibfield  {title} {\bibinfo {title} {Compressive imaging through a multimode fiber},\ }\href@noop {} {\bibfield  {journal} {\bibinfo  {journal} {Opt. Lett.}\ }\textbf {\bibinfo {volume} {43}},\ \bibinfo {pages} {5427} (\bibinfo {year} {2018})}\BibitemShut {NoStop}%
\bibitem [{\citenamefont {Amitonova}\ and\ \citenamefont {de~Boer}(2020)}]{amitonova2020endo}%
  \BibitemOpen
  \bibfield  {author} {\bibinfo {author} {\bibfnamefont {L.~V.}\ \bibnamefont {Amitonova}}\ and\ \bibinfo {author} {\bibfnamefont {J.~F.}\ \bibnamefont {de~Boer}},\ }\bibfield  {title} {\bibinfo {title} {Endo-microscopy beyond the abbe and nyquist limits},\ }\href@noop {} {\bibfield  {journal} {\bibinfo  {journal} {Light Sci. Appl.}\ }\textbf {\bibinfo {volume} {9}},\ \bibinfo {pages} {81} (\bibinfo {year} {2020})}\BibitemShut {NoStop}%
\bibitem [{\citenamefont {{\v{C}}i{\v{z}}m{\'a}r}\ and\ \citenamefont {Dholakia}(2011)}]{vcivzmar2011shaping}%
  \BibitemOpen
  \bibfield  {author} {\bibinfo {author} {\bibfnamefont {T.}~\bibnamefont {{\v{C}}i{\v{z}}m{\'a}r}}\ and\ \bibinfo {author} {\bibfnamefont {K.}~\bibnamefont {Dholakia}},\ }\bibfield  {title} {\bibinfo {title} {Shaping the light transmission through a multimode optical fibre: complex transformation analysis and applications in biophotonics},\ }\href@noop {} {\bibfield  {journal} {\bibinfo  {journal} {Opt. Express}\ }\textbf {\bibinfo {volume} {19}},\ \bibinfo {pages} {18871} (\bibinfo {year} {2011})}\BibitemShut {NoStop}%
\bibitem [{\citenamefont {Leite}\ \emph {et~al.}(2018)\citenamefont {Leite}, \citenamefont {Turtaev}, \citenamefont {Jiang}, \citenamefont {{\v{S}}iler}, \citenamefont {Cuschieri}, \citenamefont {Russell},\ and\ \citenamefont {{\v{C}}i{\v{z}}m{\'a}r}}]{leite2018three}%
  \BibitemOpen
  \bibfield  {author} {\bibinfo {author} {\bibfnamefont {I.~T.}\ \bibnamefont {Leite}}, \bibinfo {author} {\bibfnamefont {S.}~\bibnamefont {Turtaev}}, \bibinfo {author} {\bibfnamefont {X.}~\bibnamefont {Jiang}}, \bibinfo {author} {\bibfnamefont {M.}~\bibnamefont {{\v{S}}iler}}, \bibinfo {author} {\bibfnamefont {A.}~\bibnamefont {Cuschieri}}, \bibinfo {author} {\bibfnamefont {P.~S.~J.}\ \bibnamefont {Russell}},\ and\ \bibinfo {author} {\bibfnamefont {T.}~\bibnamefont {{\v{C}}i{\v{z}}m{\'a}r}},\ }\bibfield  {title} {\bibinfo {title} {Three-dimensional holographic optical manipulation through a high-numerical-aperture soft-glass multimode fibre},\ }\href@noop {} {\bibfield  {journal} {\bibinfo  {journal} {Nat. Photonics}\ }\textbf {\bibinfo {volume} {12}},\ \bibinfo {pages} {33} (\bibinfo {year} {2018})}\BibitemShut {NoStop}%
\bibitem [{\citenamefont {Li}\ \emph {et~al.}(2021{\natexlab{a}})\citenamefont {Li}, \citenamefont {Chen}, \citenamefont {Zakharian}, \citenamefont {Hurley}, \citenamefont {Stone},\ and\ \citenamefont {Li}}]{li2021large}%
  \BibitemOpen
  \bibfield  {author} {\bibinfo {author} {\bibfnamefont {K.}~\bibnamefont {Li}}, \bibinfo {author} {\bibfnamefont {X.}~\bibnamefont {Chen}}, \bibinfo {author} {\bibfnamefont {A.~R.}\ \bibnamefont {Zakharian}}, \bibinfo {author} {\bibfnamefont {J.~E.}\ \bibnamefont {Hurley}}, \bibinfo {author} {\bibfnamefont {J.~S.}\ \bibnamefont {Stone}},\ and\ \bibinfo {author} {\bibfnamefont {M.-J.}\ \bibnamefont {Li}},\ }\bibfield  {title} {\bibinfo {title} {Large core multimode fiber with high bandwidth and high connector tolerance for broadband short distance communications},\ }\href@noop {} {\bibfield  {journal} {\bibinfo  {journal} {APL Photonics}\ }\textbf {\bibinfo {volume} {6}} (\bibinfo {year} {2021}{\natexlab{a}})}\BibitemShut {NoStop}%
\bibitem [{\citenamefont {Kakkava}\ \emph {et~al.}(2019)\citenamefont {Kakkava}, \citenamefont {Romito}, \citenamefont {Conkey}, \citenamefont {Loterie}, \citenamefont {Stankovic}, \citenamefont {Moser},\ and\ \citenamefont {Psaltis}}]{kakkava2019selective}%
  \BibitemOpen
  \bibfield  {author} {\bibinfo {author} {\bibfnamefont {E.}~\bibnamefont {Kakkava}}, \bibinfo {author} {\bibfnamefont {M.}~\bibnamefont {Romito}}, \bibinfo {author} {\bibfnamefont {D.~B.}\ \bibnamefont {Conkey}}, \bibinfo {author} {\bibfnamefont {D.}~\bibnamefont {Loterie}}, \bibinfo {author} {\bibfnamefont {K.~M.}\ \bibnamefont {Stankovic}}, \bibinfo {author} {\bibfnamefont {C.}~\bibnamefont {Moser}},\ and\ \bibinfo {author} {\bibfnamefont {D.}~\bibnamefont {Psaltis}},\ }\bibfield  {title} {\bibinfo {title} {Selective femtosecond laser ablation via two-photon fluorescence imaging through a multimode fiber},\ }\href@noop {} {\bibfield  {journal} {\bibinfo  {journal} {Biomed. Opt. Express}\ }\textbf {\bibinfo {volume} {10}},\ \bibinfo {pages} {423} (\bibinfo {year} {2019})}\BibitemShut {NoStop}%
\bibitem [{\citenamefont {Dawson}\ \emph {et~al.}(2008)\citenamefont {Dawson}, \citenamefont {Messerly}, \citenamefont {Beach}, \citenamefont {Shverdin}, \citenamefont {Stappaerts}, \citenamefont {Sridharan}, \citenamefont {Pax}, \citenamefont {Heebner}, \citenamefont {Siders},\ and\ \citenamefont {Barty}}]{dawson2008analysis}%
  \BibitemOpen
  \bibfield  {author} {\bibinfo {author} {\bibfnamefont {J.~W.}\ \bibnamefont {Dawson}}, \bibinfo {author} {\bibfnamefont {M.~J.}\ \bibnamefont {Messerly}}, \bibinfo {author} {\bibfnamefont {R.~J.}\ \bibnamefont {Beach}}, \bibinfo {author} {\bibfnamefont {M.~Y.}\ \bibnamefont {Shverdin}}, \bibinfo {author} {\bibfnamefont {E.~A.}\ \bibnamefont {Stappaerts}}, \bibinfo {author} {\bibfnamefont {A.~K.}\ \bibnamefont {Sridharan}}, \bibinfo {author} {\bibfnamefont {P.~H.}\ \bibnamefont {Pax}}, \bibinfo {author} {\bibfnamefont {J.~E.}\ \bibnamefont {Heebner}}, \bibinfo {author} {\bibfnamefont {C.~W.}\ \bibnamefont {Siders}},\ and\ \bibinfo {author} {\bibfnamefont {C.}~\bibnamefont {Barty}},\ }\bibfield  {title} {\bibinfo {title} {Analysis of the scalability of diffraction-limited fiber lasers and amplifiers to high average power},\ }\href@noop {} {\bibfield  {journal} {\bibinfo  {journal} {Opt. Express}\ }\textbf {\bibinfo {volume} {16}},\ \bibinfo {pages} {13240} (\bibinfo {year} {2008})}\BibitemShut {NoStop}%
\bibitem [{\citenamefont {Richardson}\ \emph {et~al.}(2010)\citenamefont {Richardson}, \citenamefont {Nilsson},\ and\ \citenamefont {Clarkson}}]{richardson2010high}%
  \BibitemOpen
  \bibfield  {author} {\bibinfo {author} {\bibfnamefont {D.~J.}\ \bibnamefont {Richardson}}, \bibinfo {author} {\bibfnamefont {J.}~\bibnamefont {Nilsson}},\ and\ \bibinfo {author} {\bibfnamefont {W.~A.}\ \bibnamefont {Clarkson}},\ }\bibfield  {title} {\bibinfo {title} {High power fiber lasers: current status and future perspectives},\ }\href@noop {} {\bibfield  {journal} {\bibinfo  {journal} {J. Opt. Soc. Am. B}\ }\textbf {\bibinfo {volume} {27}},\ \bibinfo {pages} {B63} (\bibinfo {year} {2010})}\BibitemShut {NoStop}%
\bibitem [{\citenamefont {Zervas}\ and\ \citenamefont {Codemard}(2014)}]{zervas2014high}%
  \BibitemOpen
  \bibfield  {author} {\bibinfo {author} {\bibfnamefont {M.~N.}\ \bibnamefont {Zervas}}\ and\ \bibinfo {author} {\bibfnamefont {C.~A.}\ \bibnamefont {Codemard}},\ }\bibfield  {title} {\bibinfo {title} {High power fiber lasers: a review},\ }\href@noop {} {\bibfield  {journal} {\bibinfo  {journal} {IEEE J. Sel. Top. Quantum Electron.}\ }\textbf {\bibinfo {volume} {20}},\ \bibinfo {pages} {219} (\bibinfo {year} {2014})}\BibitemShut {NoStop}%
\bibitem [{\citenamefont {Fu}\ \emph {et~al.}(2017)\citenamefont {Fu}, \citenamefont {Shi}, \citenamefont {Feng}, \citenamefont {Zhang}, \citenamefont {Yang}, \citenamefont {Xu}, \citenamefont {Zhu}, \citenamefont {Norwood},\ and\ \citenamefont {Peyghambarian}}]{fu2017review}%
  \BibitemOpen
  \bibfield  {author} {\bibinfo {author} {\bibfnamefont {S.}~\bibnamefont {Fu}}, \bibinfo {author} {\bibfnamefont {W.}~\bibnamefont {Shi}}, \bibinfo {author} {\bibfnamefont {Y.}~\bibnamefont {Feng}}, \bibinfo {author} {\bibfnamefont {L.}~\bibnamefont {Zhang}}, \bibinfo {author} {\bibfnamefont {Z.}~\bibnamefont {Yang}}, \bibinfo {author} {\bibfnamefont {S.}~\bibnamefont {Xu}}, \bibinfo {author} {\bibfnamefont {X.}~\bibnamefont {Zhu}}, \bibinfo {author} {\bibfnamefont {R.~A.}\ \bibnamefont {Norwood}},\ and\ \bibinfo {author} {\bibfnamefont {N.}~\bibnamefont {Peyghambarian}},\ }\bibfield  {title} {\bibinfo {title} {Review of recent progress on single-frequency fiber lasers},\ }\href@noop {} {\bibfield  {journal} {\bibinfo  {journal} {J. Opt. Soc. Am. B}\ }\textbf {\bibinfo {volume} {34}},\ \bibinfo {pages} {A49} (\bibinfo {year} {2017})}\BibitemShut {NoStop}%
\bibitem [{\citenamefont {Di~Leonardo}\ and\ \citenamefont {Bianchi}(2011)}]{di2011hologram}%
  \BibitemOpen
  \bibfield  {author} {\bibinfo {author} {\bibfnamefont {R.}~\bibnamefont {Di~Leonardo}}\ and\ \bibinfo {author} {\bibfnamefont {S.}~\bibnamefont {Bianchi}},\ }\bibfield  {title} {\bibinfo {title} {Hologram transmission through multi-mode optical fibers},\ }\href@noop {} {\bibfield  {journal} {\bibinfo  {journal} {Opt. Express}\ }\textbf {\bibinfo {volume} {19}},\ \bibinfo {pages} {247} (\bibinfo {year} {2011})}\BibitemShut {NoStop}%
\bibitem [{\citenamefont {Mosk}\ \emph {et~al.}(2012)\citenamefont {Mosk}, \citenamefont {Lagendijk}, \citenamefont {Lerosey},\ and\ \citenamefont {Fink}}]{2012_Mosk_NP_Review}%
  \BibitemOpen
  \bibfield  {author} {\bibinfo {author} {\bibfnamefont {A.~P.}\ \bibnamefont {Mosk}}, \bibinfo {author} {\bibfnamefont {A.}~\bibnamefont {Lagendijk}}, \bibinfo {author} {\bibfnamefont {G.}~\bibnamefont {Lerosey}},\ and\ \bibinfo {author} {\bibfnamefont {M.}~\bibnamefont {Fink}},\ }\bibfield  {title} {\bibinfo {title} {Controlling waves in space and time for imaging and focusing in complex media},\ }\href@noop {} {\bibfield  {journal} {\bibinfo  {journal} {Nat. Photonics}\ }\textbf {\bibinfo {volume} {6}},\ \bibinfo {pages} {283} (\bibinfo {year} {2012})}\BibitemShut {NoStop}%
\bibitem [{\citenamefont {Horstmeyer}\ \emph {et~al.}(2015)\citenamefont {Horstmeyer}, \citenamefont {Ruan},\ and\ \citenamefont {Yang}}]{2015_Horstmeyer_NP_Review}%
  \BibitemOpen
  \bibfield  {author} {\bibinfo {author} {\bibfnamefont {R.}~\bibnamefont {Horstmeyer}}, \bibinfo {author} {\bibfnamefont {H.}~\bibnamefont {Ruan}},\ and\ \bibinfo {author} {\bibfnamefont {C.}~\bibnamefont {Yang}},\ }\bibfield  {title} {\bibinfo {title} {Guidestar-assisted wavefront-shaping methods for focusing light into biological tissue},\ }\href@noop {} {\bibfield  {journal} {\bibinfo  {journal} {Nat. Photonics}\ }\textbf {\bibinfo {volume} {9}},\ \bibinfo {pages} {563} (\bibinfo {year} {2015})}\BibitemShut {NoStop}%
\bibitem [{\citenamefont {Vellekoop}(2015)}]{vellekoop2015feedback}%
  \BibitemOpen
  \bibfield  {author} {\bibinfo {author} {\bibfnamefont {I.~M.}\ \bibnamefont {Vellekoop}},\ }\bibfield  {title} {\bibinfo {title} {Feedback-based wavefront shaping},\ }\href@noop {} {\bibfield  {journal} {\bibinfo  {journal} {Opt. Express}\ }\textbf {\bibinfo {volume} {23}},\ \bibinfo {pages} {12189} (\bibinfo {year} {2015})}\BibitemShut {NoStop}%
\bibitem [{\citenamefont {Rotter}\ and\ \citenamefont {Gigan}(2017)}]{2017_RotterR}%
  \BibitemOpen
  \bibfield  {author} {\bibinfo {author} {\bibfnamefont {S.}~\bibnamefont {Rotter}}\ and\ \bibinfo {author} {\bibfnamefont {S.}~\bibnamefont {Gigan}},\ }\bibfield  {title} {\bibinfo {title} {Light fields in complex media: mesoscopic scattering meets wave control},\ }\href@noop {} {\bibfield  {journal} {\bibinfo  {journal} {Rev. Mod. Phys.}\ }\textbf {\bibinfo {volume} {89}},\ \bibinfo {pages} {015005} (\bibinfo {year} {2017})}\BibitemShut {NoStop}%
\bibitem [{\citenamefont {Tzang}\ \emph {et~al.}(2018)\citenamefont {Tzang}, \citenamefont {Caravaca-Aguirre}, \citenamefont {Wagner},\ and\ \citenamefont {Piestun}}]{tzang2018adaptive}%
  \BibitemOpen
  \bibfield  {author} {\bibinfo {author} {\bibfnamefont {O.}~\bibnamefont {Tzang}}, \bibinfo {author} {\bibfnamefont {A.~M.}\ \bibnamefont {Caravaca-Aguirre}}, \bibinfo {author} {\bibfnamefont {K.}~\bibnamefont {Wagner}},\ and\ \bibinfo {author} {\bibfnamefont {R.}~\bibnamefont {Piestun}},\ }\bibfield  {title} {\bibinfo {title} {Adaptive wavefront shaping for controlling nonlinear multimode interactions in optical fibres},\ }\href@noop {} {\bibfield  {journal} {\bibinfo  {journal} {Nat. Photonics}\ }\textbf {\bibinfo {volume} {12}},\ \bibinfo {pages} {368} (\bibinfo {year} {2018})}\BibitemShut {NoStop}%
\bibitem [{\citenamefont {Gigan}\ \emph {et~al.}(2022)\citenamefont {Gigan}, \citenamefont {Katz}, \citenamefont {De~Aguiar}, \citenamefont {Andresen}, \citenamefont {Aubry}, \citenamefont {Bertolotti}, \citenamefont {Bossy}, \citenamefont {Bouchet}, \citenamefont {Brake}, \citenamefont {Brasselet} \emph {et~al.}}]{gigan2022roadmap}%
  \BibitemOpen
  \bibfield  {author} {\bibinfo {author} {\bibfnamefont {S.}~\bibnamefont {Gigan}}, \bibinfo {author} {\bibfnamefont {O.}~\bibnamefont {Katz}}, \bibinfo {author} {\bibfnamefont {H.~B.}\ \bibnamefont {De~Aguiar}}, \bibinfo {author} {\bibfnamefont {E.~R.}\ \bibnamefont {Andresen}}, \bibinfo {author} {\bibfnamefont {A.}~\bibnamefont {Aubry}}, \bibinfo {author} {\bibfnamefont {J.}~\bibnamefont {Bertolotti}}, \bibinfo {author} {\bibfnamefont {E.}~\bibnamefont {Bossy}}, \bibinfo {author} {\bibfnamefont {D.}~\bibnamefont {Bouchet}}, \bibinfo {author} {\bibfnamefont {J.}~\bibnamefont {Brake}}, \bibinfo {author} {\bibfnamefont {S.}~\bibnamefont {Brasselet}}, \emph {et~al.},\ }\bibfield  {title} {\bibinfo {title} {Roadmap on wavefront shaping and deep imaging in complex media},\ }\href@noop {} {\bibfield  {journal} {\bibinfo  {journal} {J. Phys. Photonics}\ }\textbf {\bibinfo {volume} {4}},\ \bibinfo {pages} {042501} (\bibinfo {year} {2022})}\BibitemShut {NoStop}%
\bibitem [{\citenamefont {Cao}\ \emph {et~al.}(2022)\citenamefont {Cao}, \citenamefont {Mosk},\ and\ \citenamefont {Rotter}}]{cao2022shaping}%
  \BibitemOpen
  \bibfield  {author} {\bibinfo {author} {\bibfnamefont {H.}~\bibnamefont {Cao}}, \bibinfo {author} {\bibfnamefont {A.~P.}\ \bibnamefont {Mosk}},\ and\ \bibinfo {author} {\bibfnamefont {S.}~\bibnamefont {Rotter}},\ }\bibfield  {title} {\bibinfo {title} {Shaping the propagation of light in complex media},\ }\href@noop {} {\bibfield  {journal} {\bibinfo  {journal} {Nat. Phys.}\ }\textbf {\bibinfo {volume} {18}},\ \bibinfo {pages} {994} (\bibinfo {year} {2022})}\BibitemShut {NoStop}%
\bibitem [{\citenamefont {Cao}\ \emph {et~al.}(2023)\citenamefont {Cao}, \citenamefont {{\v{C}}i{\v{z}}m{\'a}r}, \citenamefont {Turtaev}, \citenamefont {Tyc},\ and\ \citenamefont {Rotter}}]{cao2023controlling}%
  \BibitemOpen
  \bibfield  {author} {\bibinfo {author} {\bibfnamefont {H.}~\bibnamefont {Cao}}, \bibinfo {author} {\bibfnamefont {T.}~\bibnamefont {{\v{C}}i{\v{z}}m{\'a}r}}, \bibinfo {author} {\bibfnamefont {S.}~\bibnamefont {Turtaev}}, \bibinfo {author} {\bibfnamefont {T.}~\bibnamefont {Tyc}},\ and\ \bibinfo {author} {\bibfnamefont {S.}~\bibnamefont {Rotter}},\ }\bibfield  {title} {\bibinfo {title} {Controlling light propagation in multimode fibers for imaging, spectroscopy, and beyond},\ }\href@noop {} {\bibfield  {journal} {\bibinfo  {journal} {Adv. Opt. Photonics}\ }\textbf {\bibinfo {volume} {15}},\ \bibinfo {pages} {524} (\bibinfo {year} {2023})}\BibitemShut {NoStop}%
\bibitem [{\citenamefont {Chen}\ \emph {et~al.}(2023)\citenamefont {Chen}, \citenamefont {Nguyen}, \citenamefont {Wisal}, \citenamefont {Wei}, \citenamefont {Warren-Smith}, \citenamefont {Henderson-Sapir}, \citenamefont {Schartner}, \citenamefont {Ahmadi}, \citenamefont {Ebendorff-Heidepriem}, \citenamefont {Stone}, \citenamefont {Ottaway},\ and\ \citenamefont {Cao}}]{Chen2023}%
  \BibitemOpen
  \bibfield  {author} {\bibinfo {author} {\bibfnamefont {C.-W.}\ \bibnamefont {Chen}}, \bibinfo {author} {\bibfnamefont {L.~V.}\ \bibnamefont {Nguyen}}, \bibinfo {author} {\bibfnamefont {K.}~\bibnamefont {Wisal}}, \bibinfo {author} {\bibfnamefont {S.}~\bibnamefont {Wei}}, \bibinfo {author} {\bibfnamefont {S.~C.}\ \bibnamefont {Warren-Smith}}, \bibinfo {author} {\bibfnamefont {O.}~\bibnamefont {Henderson-Sapir}}, \bibinfo {author} {\bibfnamefont {E.~P.}\ \bibnamefont {Schartner}}, \bibinfo {author} {\bibfnamefont {P.}~\bibnamefont {Ahmadi}}, \bibinfo {author} {\bibfnamefont {H.}~\bibnamefont {Ebendorff-Heidepriem}}, \bibinfo {author} {\bibfnamefont {A.~D.}\ \bibnamefont {Stone}}, \bibinfo {author} {\bibfnamefont {D.~J.}\ \bibnamefont {Ottaway}},\ and\ \bibinfo {author} {\bibfnamefont {H.}~\bibnamefont {Cao}},\ }\bibfield  {title} {\bibinfo {title} {Mitigating stimulated brillouin scattering in multimode fibers with focused output via wavefront shaping},\ }\href@noop {} {\bibfield  {journal} {\bibinfo
  {journal} {Nat. Commun.}\ }\textbf {\bibinfo {volume} {14}} (\bibinfo {year} {2023})}\BibitemShut {NoStop}%
\bibitem [{\citenamefont {Wisal}\ \emph {et~al.}(2024{\natexlab{a}})\citenamefont {Wisal}, \citenamefont {Chen}, \citenamefont {Kuang}, \citenamefont {Miller}, \citenamefont {Cao},\ and\ \citenamefont {Stone}}]{wisal2024optimal}%
  \BibitemOpen
  \bibfield  {author} {\bibinfo {author} {\bibfnamefont {K.}~\bibnamefont {Wisal}}, \bibinfo {author} {\bibfnamefont {C.-W.}\ \bibnamefont {Chen}}, \bibinfo {author} {\bibfnamefont {Z.}~\bibnamefont {Kuang}}, \bibinfo {author} {\bibfnamefont {O.~D.}\ \bibnamefont {Miller}}, \bibinfo {author} {\bibfnamefont {H.}~\bibnamefont {Cao}},\ and\ \bibinfo {author} {\bibfnamefont {A.~D.}\ \bibnamefont {Stone}},\ }\bibfield  {title} {\bibinfo {title} {Optimal input excitations for suppressing nonlinear instabilities in multimode fibers},\ }\href@noop {} {\bibfield  {journal} {\bibinfo  {journal} {Optica}\ }\textbf {\bibinfo {volume} {11}},\ \bibinfo {pages} {1663} (\bibinfo {year} {2024}{\natexlab{a}})}\BibitemShut {NoStop}%
\bibitem [{\citenamefont {Popoff}\ \emph {et~al.}(2010)\citenamefont {Popoff}, \citenamefont {Lerosey}, \citenamefont {Carminati}, \citenamefont {Fink}, \citenamefont {Boccara},\ and\ \citenamefont {Gigan}}]{2010_Popoff_PRL}%
  \BibitemOpen
  \bibfield  {author} {\bibinfo {author} {\bibfnamefont {S.~M.}\ \bibnamefont {Popoff}}, \bibinfo {author} {\bibfnamefont {G.}~\bibnamefont {Lerosey}}, \bibinfo {author} {\bibfnamefont {R.}~\bibnamefont {Carminati}}, \bibinfo {author} {\bibfnamefont {M.}~\bibnamefont {Fink}}, \bibinfo {author} {\bibfnamefont {A.~C.}\ \bibnamefont {Boccara}},\ and\ \bibinfo {author} {\bibfnamefont {S.}~\bibnamefont {Gigan}},\ }\bibfield  {title} {\bibinfo {title} {Measuring the transmission matrix in optics: an approach to the study and control of light propagation in disordered media},\ }\href@noop {} {\bibfield  {journal} {\bibinfo  {journal} {Phys. Rev. Lett.}\ }\textbf {\bibinfo {volume} {104}},\ \bibinfo {pages} {100601} (\bibinfo {year} {2010})}\BibitemShut {NoStop}%
\bibitem [{\citenamefont {Kim}\ \emph {et~al.}(2015)\citenamefont {Kim}, \citenamefont {Choi}, \citenamefont {Choi}, \citenamefont {Yoon},\ and\ \citenamefont {Choi}}]{2015_Choi_OptExpress_R}%
  \BibitemOpen
  \bibfield  {author} {\bibinfo {author} {\bibfnamefont {M.}~\bibnamefont {Kim}}, \bibinfo {author} {\bibfnamefont {W.}~\bibnamefont {Choi}}, \bibinfo {author} {\bibfnamefont {Y.}~\bibnamefont {Choi}}, \bibinfo {author} {\bibfnamefont {C.}~\bibnamefont {Yoon}},\ and\ \bibinfo {author} {\bibfnamefont {W.}~\bibnamefont {Choi}},\ }\bibfield  {title} {\bibinfo {title} {Transmission matrix of a scattering medium and its applications in biophotonics},\ }\href@noop {} {\bibfield  {journal} {\bibinfo  {journal} {Opt. Express}\ }\textbf {\bibinfo {volume} {23}},\ \bibinfo {pages} {12648} (\bibinfo {year} {2015})}\BibitemShut {NoStop}%
\bibitem [{\citenamefont {Li}\ \emph {et~al.}(2021{\natexlab{b}})\citenamefont {Li}, \citenamefont {Horsley}, \citenamefont {Tyc}, \citenamefont {Čižmár},\ and\ \citenamefont {Phillips}}]{2021_Li_NC}%
  \BibitemOpen
  \bibfield  {author} {\bibinfo {author} {\bibfnamefont {S.}~\bibnamefont {Li}}, \bibinfo {author} {\bibfnamefont {S.~A.~R.}\ \bibnamefont {Horsley}}, \bibinfo {author} {\bibfnamefont {T.}~\bibnamefont {Tyc}}, \bibinfo {author} {\bibfnamefont {T.}~\bibnamefont {Čižmár}},\ and\ \bibinfo {author} {\bibfnamefont {D.~B.}\ \bibnamefont {Phillips}},\ }\bibfield  {title} {\bibinfo {title} {Memory effect assisted imaging through multimode optical fibres},\ }\href@noop {} {\bibfield  {journal} {\bibinfo  {journal} {Nat. Commun.}\ }\textbf {\bibinfo {volume} {12}},\ \bibinfo {pages} {3751} (\bibinfo {year} {2021}{\natexlab{b}})}\BibitemShut {NoStop}%
\bibitem [{\citenamefont {Bender}\ \emph {et~al.}(2023)\citenamefont {Bender}, \citenamefont {Haig}, \citenamefont {Christodoulides},\ and\ \citenamefont {Wise}}]{bender2023spectral}%
  \BibitemOpen
  \bibfield  {author} {\bibinfo {author} {\bibfnamefont {N.}~\bibnamefont {Bender}}, \bibinfo {author} {\bibfnamefont {H.}~\bibnamefont {Haig}}, \bibinfo {author} {\bibfnamefont {D.~N.}\ \bibnamefont {Christodoulides}},\ and\ \bibinfo {author} {\bibfnamefont {F.~W.}\ \bibnamefont {Wise}},\ }\bibfield  {title} {\bibinfo {title} {Spectral speckle customization},\ }\href@noop {} {\bibfield  {journal} {\bibinfo  {journal} {Optica}\ }\textbf {\bibinfo {volume} {10}},\ \bibinfo {pages} {1260} (\bibinfo {year} {2023})}\BibitemShut {NoStop}%
\bibitem [{\citenamefont {Vellekoop}\ and\ \citenamefont {Mosk}(2007)}]{2007_Vellekoop_OL}%
  \BibitemOpen
  \bibfield  {author} {\bibinfo {author} {\bibfnamefont {I.~M.}\ \bibnamefont {Vellekoop}}\ and\ \bibinfo {author} {\bibfnamefont {A.~P.}\ \bibnamefont {Mosk}},\ }\bibfield  {title} {\bibinfo {title} {Focusing coherent light through opaque strongly scattering media},\ }\href@noop {} {\bibfield  {journal} {\bibinfo  {journal} {Opt. Lett.}\ }\textbf {\bibinfo {volume} {32}},\ \bibinfo {pages} {2309} (\bibinfo {year} {2007})}\BibitemShut {NoStop}%
\bibitem [{\citenamefont {Vellekoop}\ \emph {et~al.}(2010)\citenamefont {Vellekoop}, \citenamefont {Lagendijk},\ and\ \citenamefont {Mosk}}]{vellekoop2010exploiting}%
  \BibitemOpen
  \bibfield  {author} {\bibinfo {author} {\bibfnamefont {I.~M.}\ \bibnamefont {Vellekoop}}, \bibinfo {author} {\bibfnamefont {A.}~\bibnamefont {Lagendijk}},\ and\ \bibinfo {author} {\bibfnamefont {A.}~\bibnamefont {Mosk}},\ }\bibfield  {title} {\bibinfo {title} {Exploiting disorder for perfect focusing},\ }\href@noop {} {\bibfield  {journal} {\bibinfo  {journal} {Nat. Photonics}\ }\textbf {\bibinfo {volume} {4}},\ \bibinfo {pages} {320} (\bibinfo {year} {2010})}\BibitemShut {NoStop}%
\bibitem [{\citenamefont {Caravaca-Aguirre}\ \emph {et~al.}(2013)\citenamefont {Caravaca-Aguirre}, \citenamefont {Niv}, \citenamefont {Conkey},\ and\ \citenamefont {Piestun}}]{2013_Caravaca-Aguirre_OE}%
  \BibitemOpen
  \bibfield  {author} {\bibinfo {author} {\bibfnamefont {A.~M.}\ \bibnamefont {Caravaca-Aguirre}}, \bibinfo {author} {\bibfnamefont {E.}~\bibnamefont {Niv}}, \bibinfo {author} {\bibfnamefont {D.~B.}\ \bibnamefont {Conkey}},\ and\ \bibinfo {author} {\bibfnamefont {R.}~\bibnamefont {Piestun}},\ }\bibfield  {title} {\bibinfo {title} {Real-time resilient focusing through a bending multimode fiber},\ }\href@noop {} {\bibfield  {journal} {\bibinfo  {journal} {Opt. Express}\ }\textbf {\bibinfo {volume} {21}},\ \bibinfo {pages} {12881} (\bibinfo {year} {2013})}\BibitemShut {NoStop}%
\bibitem [{\citenamefont {Dr{\'e}meau}\ \emph {et~al.}(2015)\citenamefont {Dr{\'e}meau}, \citenamefont {Liutkus}, \citenamefont {Martina}, \citenamefont {Katz}, \citenamefont {Sch{\"u}lke}, \citenamefont {Krzakala}, \citenamefont {Gigan},\ and\ \citenamefont {Daudet}}]{dremeau2015reference}%
  \BibitemOpen
  \bibfield  {author} {\bibinfo {author} {\bibfnamefont {A.}~\bibnamefont {Dr{\'e}meau}}, \bibinfo {author} {\bibfnamefont {A.}~\bibnamefont {Liutkus}}, \bibinfo {author} {\bibfnamefont {D.}~\bibnamefont {Martina}}, \bibinfo {author} {\bibfnamefont {O.}~\bibnamefont {Katz}}, \bibinfo {author} {\bibfnamefont {C.}~\bibnamefont {Sch{\"u}lke}}, \bibinfo {author} {\bibfnamefont {F.}~\bibnamefont {Krzakala}}, \bibinfo {author} {\bibfnamefont {S.}~\bibnamefont {Gigan}},\ and\ \bibinfo {author} {\bibfnamefont {L.}~\bibnamefont {Daudet}},\ }\bibfield  {title} {\bibinfo {title} {Reference-less measurement of the transmission matrix of a highly scattering material using a {DMD} and phase retrieval techniques},\ }\href@noop {} {\bibfield  {journal} {\bibinfo  {journal} {Opt. Express}\ }\textbf {\bibinfo {volume} {23}},\ \bibinfo {pages} {11898} (\bibinfo {year} {2015})}\BibitemShut {NoStop}%
\bibitem [{\citenamefont {Amitonova}\ \emph {et~al.}(2016)\citenamefont {Amitonova}, \citenamefont {Descloux}, \citenamefont {Petschulat}, \citenamefont {Frosz}, \citenamefont {Ahmed}, \citenamefont {Babic}, \citenamefont {Jiang}, \citenamefont {Mosk}, \citenamefont {Russell},\ and\ \citenamefont {Pinkse}}]{2016_Amitonova_OL}%
  \BibitemOpen
  \bibfield  {author} {\bibinfo {author} {\bibfnamefont {L.~V.}\ \bibnamefont {Amitonova}}, \bibinfo {author} {\bibfnamefont {A.}~\bibnamefont {Descloux}}, \bibinfo {author} {\bibfnamefont {J.}~\bibnamefont {Petschulat}}, \bibinfo {author} {\bibfnamefont {M.~H.}\ \bibnamefont {Frosz}}, \bibinfo {author} {\bibfnamefont {G.}~\bibnamefont {Ahmed}}, \bibinfo {author} {\bibfnamefont {F.}~\bibnamefont {Babic}}, \bibinfo {author} {\bibfnamefont {X.}~\bibnamefont {Jiang}}, \bibinfo {author} {\bibfnamefont {A.~P.}\ \bibnamefont {Mosk}}, \bibinfo {author} {\bibfnamefont {P.~S.}\ \bibnamefont {Russell}},\ and\ \bibinfo {author} {\bibfnamefont {P.~W.~H.}\ \bibnamefont {Pinkse}},\ }\bibfield  {title} {\bibinfo {title} {High-resolution wavefront shaping with a photonic crystal fiber for multimode fiber imaging},\ }\href@noop {} {\bibfield  {journal} {\bibinfo  {journal} {Opt. Lett.}\ }\textbf {\bibinfo {volume} {41}},\ \bibinfo {pages} {497} (\bibinfo {year} {2016})}\BibitemShut {NoStop}%
\bibitem [{\citenamefont {Descloux}\ \emph {et~al.}(2016)\citenamefont {Descloux}, \citenamefont {Amitonova},\ and\ \citenamefont {Pinkse}}]{2016_Descloux_OE}%
  \BibitemOpen
  \bibfield  {author} {\bibinfo {author} {\bibfnamefont {A.}~\bibnamefont {Descloux}}, \bibinfo {author} {\bibfnamefont {L.~V.}\ \bibnamefont {Amitonova}},\ and\ \bibinfo {author} {\bibfnamefont {P.~W.~H.}\ \bibnamefont {Pinkse}},\ }\bibfield  {title} {\bibinfo {title} {Aberrations of the point spread function of a multimode fiber due to partial mode excitation},\ }\href@noop {} {\bibfield  {journal} {\bibinfo  {journal} {Opt. Express}\ }\textbf {\bibinfo {volume} {24}},\ \bibinfo {pages} {18501} (\bibinfo {year} {2016})}\BibitemShut {NoStop}%
\bibitem [{\citenamefont {N’Gom}\ \emph {et~al.}(2018)\citenamefont {N’Gom}, \citenamefont {Norris}, \citenamefont {Michielssen},\ and\ \citenamefont {Nadakuditi}}]{n2018mode}%
  \BibitemOpen
  \bibfield  {author} {\bibinfo {author} {\bibfnamefont {M.}~\bibnamefont {N’Gom}}, \bibinfo {author} {\bibfnamefont {T.~B.}\ \bibnamefont {Norris}}, \bibinfo {author} {\bibfnamefont {E.}~\bibnamefont {Michielssen}},\ and\ \bibinfo {author} {\bibfnamefont {R.~R.}\ \bibnamefont {Nadakuditi}},\ }\bibfield  {title} {\bibinfo {title} {Mode control in a multimode fiber through acquiring its transmission matrix from a reference-less optical system},\ }\href@noop {} {\bibfield  {journal} {\bibinfo  {journal} {Opt. Lett.}\ }\textbf {\bibinfo {volume} {43}},\ \bibinfo {pages} {419} (\bibinfo {year} {2018})}\BibitemShut {NoStop}%
\bibitem [{\citenamefont {Nam}\ and\ \citenamefont {Park}(2020)}]{nam2020increasing}%
  \BibitemOpen
  \bibfield  {author} {\bibinfo {author} {\bibfnamefont {K.}~\bibnamefont {Nam}}\ and\ \bibinfo {author} {\bibfnamefont {J.-H.}\ \bibnamefont {Park}},\ }\bibfield  {title} {\bibinfo {title} {Increasing the enhancement factor for {DMD}-based wavefront shaping},\ }\href@noop {} {\bibfield  {journal} {\bibinfo  {journal} {Opt. Lett.}\ }\textbf {\bibinfo {volume} {45}},\ \bibinfo {pages} {3381} (\bibinfo {year} {2020})}\BibitemShut {NoStop}%
\bibitem [{\citenamefont {Lyu}\ \emph {et~al.}(2022)\citenamefont {Lyu}, \citenamefont {Osnabrugge}, \citenamefont {Pinkse},\ and\ \citenamefont {Amitonova}}]{2022_Lyu_AO}%
  \BibitemOpen
  \bibfield  {author} {\bibinfo {author} {\bibfnamefont {Z.}~\bibnamefont {Lyu}}, \bibinfo {author} {\bibfnamefont {G.}~\bibnamefont {Osnabrugge}}, \bibinfo {author} {\bibfnamefont {P.~W.~H.}\ \bibnamefont {Pinkse}},\ and\ \bibinfo {author} {\bibfnamefont {L.~V.}\ \bibnamefont {Amitonova}},\ }\bibfield  {title} {\bibinfo {title} {Focus quality in raster-scan imaging via a multimode fiber},\ }\href@noop {} {\bibfield  {journal} {\bibinfo  {journal} {Appl. Opt.}\ }\textbf {\bibinfo {volume} {61}},\ \bibinfo {pages} {4363} (\bibinfo {year} {2022})}\BibitemShut {NoStop}%
\bibitem [{\citenamefont {Gomes}\ \emph {et~al.}(2022)\citenamefont {Gomes}, \citenamefont {Turtaev}, \citenamefont {Du},\ and\ \citenamefont {{\v{C}}i{\v{z}}m{\'a}r}}]{gomes2022near}%
  \BibitemOpen
  \bibfield  {author} {\bibinfo {author} {\bibfnamefont {A.~D.}\ \bibnamefont {Gomes}}, \bibinfo {author} {\bibfnamefont {S.}~\bibnamefont {Turtaev}}, \bibinfo {author} {\bibfnamefont {Y.}~\bibnamefont {Du}},\ and\ \bibinfo {author} {\bibfnamefont {T.}~\bibnamefont {{\v{C}}i{\v{z}}m{\'a}r}},\ }\bibfield  {title} {\bibinfo {title} {Near perfect focusing through multimode fibres},\ }\href@noop {} {\bibfield  {journal} {\bibinfo  {journal} {Opt. Express}\ }\textbf {\bibinfo {volume} {30}},\ \bibinfo {pages} {10645} (\bibinfo {year} {2022})}\BibitemShut {NoStop}%
\bibitem [{\citenamefont {Lyu}\ and\ \citenamefont {Amitonova}(2024)}]{lyu2024wavefront}%
  \BibitemOpen
  \bibfield  {author} {\bibinfo {author} {\bibfnamefont {Z.}~\bibnamefont {Lyu}}\ and\ \bibinfo {author} {\bibfnamefont {L.~V.}\ \bibnamefont {Amitonova}},\ }\bibfield  {title} {\bibinfo {title} {Wavefront shaping and imaging through a multimode hollow-core fiber},\ }\href@noop {} {\bibfield  {journal} {\bibinfo  {journal} {Opt. Express}\ }\textbf {\bibinfo {volume} {32}},\ \bibinfo {pages} {37098} (\bibinfo {year} {2024})}\BibitemShut {NoStop}%
\bibitem [{\citenamefont {Hammer}\ and\ \citenamefont {Uppu}(2025)}]{hammer2025effect}%
  \BibitemOpen
  \bibfield  {author} {\bibinfo {author} {\bibfnamefont {H.~C.}\ \bibnamefont {Hammer}}\ and\ \bibinfo {author} {\bibfnamefont {R.}~\bibnamefont {Uppu}},\ }\bibfield  {title} {\bibinfo {title} {Effect of core geometry on frequency correlations and channel capacity of a multimode optical fiber},\ }\href@noop {} {\bibfield  {journal} {\bibinfo  {journal} {Adv. Photonics Res.}\ }\textbf {\bibinfo {volume} {6}},\ \bibinfo {pages} {2400156} (\bibinfo {year} {2025})}\BibitemShut {NoStop}%
\bibitem [{\citenamefont {Mastiani}\ \emph {et~al.}(2024)\citenamefont {Mastiani}, \citenamefont {Cox},\ and\ \citenamefont {Vellekoop}}]{mastiani2024practical}%
  \BibitemOpen
  \bibfield  {author} {\bibinfo {author} {\bibfnamefont {B.}~\bibnamefont {Mastiani}}, \bibinfo {author} {\bibfnamefont {D.}~\bibnamefont {Cox}},\ and\ \bibinfo {author} {\bibfnamefont {I.~M.}\ \bibnamefont {Vellekoop}},\ }\bibfield  {title} {\bibinfo {title} {Practical considerations for high-fidelity wavefront shaping experiments},\ }\href@noop {} {\bibfield  {journal} {\bibinfo  {journal} {J. Phys. Photonics}\ }\textbf {\bibinfo {volume} {6}},\ \bibinfo {pages} {033003} (\bibinfo {year} {2024})}\BibitemShut {NoStop}%
\bibitem [{\citenamefont {Vellekoop}\ and\ \citenamefont {Mosk}(2008{\natexlab{a}})}]{vellekoop2008phase}%
  \BibitemOpen
  \bibfield  {author} {\bibinfo {author} {\bibfnamefont {I.~M.}\ \bibnamefont {Vellekoop}}\ and\ \bibinfo {author} {\bibfnamefont {A.}~\bibnamefont {Mosk}},\ }\bibfield  {title} {\bibinfo {title} {Phase control algorithms for focusing light through turbid media},\ }\href@noop {} {\bibfield  {journal} {\bibinfo  {journal} {Opt. Commun.}\ }\textbf {\bibinfo {volume} {281}},\ \bibinfo {pages} {3071} (\bibinfo {year} {2008}{\natexlab{a}})}\BibitemShut {NoStop}%
\bibitem [{\citenamefont {Y{\i}lmaz}\ \emph {et~al.}(2013)\citenamefont {Y{\i}lmaz}, \citenamefont {Vos},\ and\ \citenamefont {Mosk}}]{2013_Yilmaz_BOE}%
  \BibitemOpen
  \bibfield  {author} {\bibinfo {author} {\bibfnamefont {H.}~\bibnamefont {Y{\i}lmaz}}, \bibinfo {author} {\bibfnamefont {W.~L.}\ \bibnamefont {Vos}},\ and\ \bibinfo {author} {\bibfnamefont {A.~P.}\ \bibnamefont {Mosk}},\ }\bibfield  {title} {\bibinfo {title} {Optimal control of light propagation through multiple-scattering media in the presence of noise},\ }\href@noop {} {\bibfield  {journal} {\bibinfo  {journal} {Biomed. Opt. Express}\ }\textbf {\bibinfo {volume} {4}},\ \bibinfo {pages} {1759} (\bibinfo {year} {2013})}\BibitemShut {NoStop}%
\bibitem [{\citenamefont {Mastiani}\ and\ \citenamefont {Vellekoop}(2021)}]{mastiani2021noise}%
  \BibitemOpen
  \bibfield  {author} {\bibinfo {author} {\bibfnamefont {B.}~\bibnamefont {Mastiani}}\ and\ \bibinfo {author} {\bibfnamefont {I.~M.}\ \bibnamefont {Vellekoop}},\ }\bibfield  {title} {\bibinfo {title} {Noise-tolerant wavefront shaping in a hadamard basis},\ }\href@noop {} {\bibfield  {journal} {\bibinfo  {journal} {Opt. Express}\ }\textbf {\bibinfo {volume} {29}},\ \bibinfo {pages} {17534} (\bibinfo {year} {2021})}\BibitemShut {NoStop}%
\bibitem [{\citenamefont {Resisi}\ \emph {et~al.}(2020)\citenamefont {Resisi}, \citenamefont {Viernik}, \citenamefont {Popoff},\ and\ \citenamefont {Bromberg}}]{2020_Resisi_APLP}%
  \BibitemOpen
  \bibfield  {author} {\bibinfo {author} {\bibfnamefont {S.}~\bibnamefont {Resisi}}, \bibinfo {author} {\bibfnamefont {Y.}~\bibnamefont {Viernik}}, \bibinfo {author} {\bibfnamefont {S.~M.}\ \bibnamefont {Popoff}},\ and\ \bibinfo {author} {\bibfnamefont {Y.}~\bibnamefont {Bromberg}},\ }\bibfield  {title} {\bibinfo {title} {Wavefront shaping in multimode fibers by transmission matrix engineering},\ }\href@noop {} {\bibfield  {journal} {\bibinfo  {journal} {APL Photonics}\ }\textbf {\bibinfo {volume} {5}},\ \bibinfo {pages} {036103} (\bibinfo {year} {2020})}\BibitemShut {NoStop}%
\bibitem [{\citenamefont {Wisal}\ \emph {et~al.}(2024{\natexlab{b}})\citenamefont {Wisal}, \citenamefont {Chen}, \citenamefont {Cao},\ and\ \citenamefont {Stone}}]{wisal2024theory}%
  \BibitemOpen
  \bibfield  {author} {\bibinfo {author} {\bibfnamefont {K.}~\bibnamefont {Wisal}}, \bibinfo {author} {\bibfnamefont {C.-W.}\ \bibnamefont {Chen}}, \bibinfo {author} {\bibfnamefont {H.}~\bibnamefont {Cao}},\ and\ \bibinfo {author} {\bibfnamefont {A.~D.}\ \bibnamefont {Stone}},\ }\bibfield  {title} {\bibinfo {title} {Theory of transverse mode instability in fiber amplifiers with multimode excitations},\ }\href@noop {} {\bibfield  {journal} {\bibinfo  {journal} {APL Photonics}\ }\textbf {\bibinfo {volume} {9}} (\bibinfo {year} {2024}{\natexlab{b}})}\BibitemShut {NoStop}%
\bibitem [{\citenamefont {Rothe}\ \emph {et~al.}(2025{\natexlab{a}})\citenamefont {Rothe}, \citenamefont {Wisal}, \citenamefont {Chen}, \citenamefont {Ercan}, \citenamefont {Jesacher}, \citenamefont {Stone},\ and\ \citenamefont {Cao}}]{rothe2025output}%
  \BibitemOpen
  \bibfield  {author} {\bibinfo {author} {\bibfnamefont {S.}~\bibnamefont {Rothe}}, \bibinfo {author} {\bibfnamefont {K.}~\bibnamefont {Wisal}}, \bibinfo {author} {\bibfnamefont {C.-W.}\ \bibnamefont {Chen}}, \bibinfo {author} {\bibfnamefont {M.}~\bibnamefont {Ercan}}, \bibinfo {author} {\bibfnamefont {A.}~\bibnamefont {Jesacher}}, \bibinfo {author} {\bibfnamefont {A.~D.}\ \bibnamefont {Stone}},\ and\ \bibinfo {author} {\bibfnamefont {H.}~\bibnamefont {Cao}},\ }\bibfield  {title} {\bibinfo {title} {Output beam shaping of a multimode fiber amplifier},\ }\href@noop {} {\bibfield  {journal} {\bibinfo  {journal} {Opt. Commun.}\ }\textbf {\bibinfo {volume} {577}},\ \bibinfo {pages} {131405} (\bibinfo {year} {2025}{\natexlab{a}})}\BibitemShut {NoStop}%
\bibitem [{\citenamefont {Rothe}\ \emph {et~al.}(2025{\natexlab{b}})\citenamefont {Rothe}, \citenamefont {Chen}, \citenamefont {Ahmadi}, \citenamefont {Wisal}, \citenamefont {Ercan}, \citenamefont {Lee}, \citenamefont {Vigne}, \citenamefont {Stone},\ and\ \citenamefont {Cao}}]{2025_Rothe_Science}%
  \BibitemOpen
  \bibfield  {author} {\bibinfo {author} {\bibfnamefont {S.}~\bibnamefont {Rothe}}, \bibinfo {author} {\bibfnamefont {C.-W.}\ \bibnamefont {Chen}}, \bibinfo {author} {\bibfnamefont {P.}~\bibnamefont {Ahmadi}}, \bibinfo {author} {\bibfnamefont {K.}~\bibnamefont {Wisal}}, \bibinfo {author} {\bibfnamefont {M.}~\bibnamefont {Ercan}}, \bibinfo {author} {\bibfnamefont {K.}~\bibnamefont {Lee}}, \bibinfo {author} {\bibfnamefont {N.}~\bibnamefont {Vigne}}, \bibinfo {author} {\bibfnamefont {A.~D.}\ \bibnamefont {Stone}},\ and\ \bibinfo {author} {\bibfnamefont {H.}~\bibnamefont {Cao}},\ }\bibfield  {title} {\bibinfo {title} {Wavefront shaping enables high-power multimode fiber amplifier with output control},\ }\href@noop {} {\bibfield  {journal} {\bibinfo  {journal} {Science}\ }\textbf {\bibinfo {volume} {390}},\ \bibinfo {pages} {173} (\bibinfo {year} {2025}{\natexlab{b}})}\BibitemShut {NoStop}%
\bibitem [{\citenamefont {van Beijnum}(2009)}]{vanBeijnum2008thesis}%
  \BibitemOpen
  \bibfield  {author} {\bibinfo {author} {\bibfnamefont {F.}~\bibnamefont {van Beijnum}},\ }\emph {\bibinfo {title} {Light takes no shortcuts}},\ \href@noop {} {Master's thesis},\ \bibinfo  {school} {Department of Science and Technology, University of Twente} (\bibinfo {year} {2009})\BibitemShut {NoStop}%
\bibitem [{\citenamefont {Hsu}\ \emph {et~al.}(2017)\citenamefont {Hsu}, \citenamefont {Liew}, \citenamefont {Goetschy}, \citenamefont {Cao},\ and\ \citenamefont {Stone}}]{Wade1}%
  \BibitemOpen
  \bibfield  {author} {\bibinfo {author} {\bibfnamefont {C.~W.}\ \bibnamefont {Hsu}}, \bibinfo {author} {\bibfnamefont {S.~F.}\ \bibnamefont {Liew}}, \bibinfo {author} {\bibfnamefont {A.}~\bibnamefont {Goetschy}}, \bibinfo {author} {\bibfnamefont {H.}~\bibnamefont {Cao}},\ and\ \bibinfo {author} {\bibfnamefont {A.~D.}\ \bibnamefont {Stone}},\ }\bibfield  {title} {\bibinfo {title} {Correlation-enhanced control of wave focusing in disordered media},\ }\href@noop {} {\bibfield  {journal} {\bibinfo  {journal} {Nat. Phys.}\ }\textbf {\bibinfo {volume} {13}},\ \bibinfo {pages} {497} (\bibinfo {year} {2017})}\BibitemShut {NoStop}%
\bibitem [{\citenamefont {Yılmaz}\ \emph {et~al.}(2019)\citenamefont {Yılmaz}, \citenamefont {Hsu}, \citenamefont {Yamilov},\ and\ \citenamefont {Cao}}]{Yilmaz2019}%
  \BibitemOpen
  \bibfield  {author} {\bibinfo {author} {\bibfnamefont {H.}~\bibnamefont {Yılmaz}}, \bibinfo {author} {\bibfnamefont {C.~W.}\ \bibnamefont {Hsu}}, \bibinfo {author} {\bibfnamefont {A.}~\bibnamefont {Yamilov}},\ and\ \bibinfo {author} {\bibfnamefont {H.}~\bibnamefont {Cao}},\ }\bibfield  {title} {\bibinfo {title} {Transverse localization of transmission eigenchannels},\ }\href@noop {} {\bibfield  {journal} {\bibinfo  {journal} {Nat. Photonics}\ }\textbf {\bibinfo {volume} {13}},\ \bibinfo {pages} {352} (\bibinfo {year} {2019})}\BibitemShut {NoStop}%
\bibitem [{\citenamefont {Y{\i}lmaz}\ \emph {et~al.}(2019)\citenamefont {Y{\i}lmaz}, \citenamefont {Hsu}, \citenamefont {Goetschy}, \citenamefont {Bittner}, \citenamefont {Rotter}, \citenamefont {Yamilov},\ and\ \citenamefont {Cao}}]{Yilmaz2019memory}%
  \BibitemOpen
  \bibfield  {author} {\bibinfo {author} {\bibfnamefont {H.}~\bibnamefont {Y{\i}lmaz}}, \bibinfo {author} {\bibfnamefont {C.~W.}\ \bibnamefont {Hsu}}, \bibinfo {author} {\bibfnamefont {A.}~\bibnamefont {Goetschy}}, \bibinfo {author} {\bibfnamefont {S.}~\bibnamefont {Bittner}}, \bibinfo {author} {\bibfnamefont {S.}~\bibnamefont {Rotter}}, \bibinfo {author} {\bibfnamefont {A.}~\bibnamefont {Yamilov}},\ and\ \bibinfo {author} {\bibfnamefont {H.}~\bibnamefont {Cao}},\ }\bibfield  {title} {\bibinfo {title} {Angular memory effect of transmission eigenchannels},\ }\href {https://doi.org/10.1103/PhysRevLett.123.203901} {\bibfield  {journal} {\bibinfo  {journal} {Phys. Rev. Lett.}\ }\textbf {\bibinfo {volume} {123}},\ \bibinfo {pages} {203901} (\bibinfo {year} {2019})}\BibitemShut {NoStop}%
\bibitem [{\citenamefont {van Putten}(2011)}]{vanPutten2011thesis}%
  \BibitemOpen
  \bibfield  {author} {\bibinfo {author} {\bibfnamefont {E.~G.}\ \bibnamefont {van Putten}},\ }\emph {\bibinfo {title} {Disorder-enhanced imaging with spatially controlled light}},\ \href@noop {} {Ph.D. thesis},\ \bibinfo  {school} {Department of Science and Technology, University of Twente} (\bibinfo {year} {2011})\BibitemShut {NoStop}%
\bibitem [{\citenamefont {Duan}\ \emph {et~al.}(2023)\citenamefont {Duan}, \citenamefont {Zhao}, \citenamefont {Yang}, \citenamefont {Deng}, \citenamefont {Huangfu}, \citenamefont {Zuo}, \citenamefont {Li},\ and\ \citenamefont {Wang}}]{duan2023modulate}%
  \BibitemOpen
  \bibfield  {author} {\bibinfo {author} {\bibfnamefont {M.}~\bibnamefont {Duan}}, \bibinfo {author} {\bibfnamefont {Y.}~\bibnamefont {Zhao}}, \bibinfo {author} {\bibfnamefont {Z.}~\bibnamefont {Yang}}, \bibinfo {author} {\bibfnamefont {X.}~\bibnamefont {Deng}}, \bibinfo {author} {\bibfnamefont {H.}~\bibnamefont {Huangfu}}, \bibinfo {author} {\bibfnamefont {H.}~\bibnamefont {Zuo}}, \bibinfo {author} {\bibfnamefont {Z.}~\bibnamefont {Li}},\ and\ \bibinfo {author} {\bibfnamefont {D.}~\bibnamefont {Wang}},\ }\bibfield  {title} {\bibinfo {title} {Modulate scattered light field with point guard algorithm},\ }\href@noop {} {\bibfield  {journal} {\bibinfo  {journal} {Opt. Commun.}\ }\textbf {\bibinfo {volume} {548}},\ \bibinfo {pages} {129832} (\bibinfo {year} {2023})}\BibitemShut {NoStop}%
\bibitem [{\citenamefont {Vellekoop}\ and\ \citenamefont {Mosk}(2008{\natexlab{b}})}]{2008_Vellekoop_PRL}%
  \BibitemOpen
  \bibfield  {author} {\bibinfo {author} {\bibfnamefont {I.~M.}\ \bibnamefont {Vellekoop}}\ and\ \bibinfo {author} {\bibfnamefont {A.~P.}\ \bibnamefont {Mosk}},\ }\bibfield  {title} {\bibinfo {title} {Universal optimal transmission of light through disordered materials},\ }\href@noop {} {\bibfield  {journal} {\bibinfo  {journal} {Phys. Rev. Lett.}\ }\textbf {\bibinfo {volume} {101}},\ \bibinfo {pages} {120601} (\bibinfo {year} {2008}{\natexlab{b}})}\BibitemShut {NoStop}%
\bibitem [{\citenamefont {Shaughnessy}\ \emph {et~al.}(2024)\citenamefont {Shaughnessy}, \citenamefont {McIntosh}, \citenamefont {Goetschy}, \citenamefont {Hsu}, \citenamefont {Bender}, \citenamefont {Y{\i}lmaz}, \citenamefont {Yamilov},\ and\ \citenamefont {Cao}}]{shaughnessy2024multiregion}%
  \BibitemOpen
  \bibfield  {author} {\bibinfo {author} {\bibfnamefont {L.}~\bibnamefont {Shaughnessy}}, \bibinfo {author} {\bibfnamefont {R.~E.}\ \bibnamefont {McIntosh}}, \bibinfo {author} {\bibfnamefont {A.}~\bibnamefont {Goetschy}}, \bibinfo {author} {\bibfnamefont {C.~W.}\ \bibnamefont {Hsu}}, \bibinfo {author} {\bibfnamefont {N.}~\bibnamefont {Bender}}, \bibinfo {author} {\bibfnamefont {H.}~\bibnamefont {Y{\i}lmaz}}, \bibinfo {author} {\bibfnamefont {A.}~\bibnamefont {Yamilov}},\ and\ \bibinfo {author} {\bibfnamefont {H.}~\bibnamefont {Cao}},\ }\bibfield  {title} {\bibinfo {title} {Multiregion light control in diffusive media via wavefront shaping},\ }\href@noop {} {\bibfield  {journal} {\bibinfo  {journal} {Phys. Rev. Lett.}\ }\textbf {\bibinfo {volume} {133}},\ \bibinfo {pages} {146901} (\bibinfo {year} {2024})}\BibitemShut {NoStop}%
\bibitem [{\citenamefont {Lee}(2023)}]{Lee2023GitHub}%
  \BibitemOpen
  \bibfield  {author} {\bibinfo {author} {\bibfnamefont {S.-Y.}\ \bibnamefont {Lee}},\ }\href@noop {} {\bibinfo {title} {{MMF}-simulation}},\ \bibinfo {howpublished} {\url{https://github.com/szuyul/MMF-simulation/releases/tag/v1.0}} (\bibinfo {year} {2023})\BibitemShut {NoStop}%
\bibitem [{\citenamefont {Jang}\ \emph {et~al.}(2017)\citenamefont {Jang}, \citenamefont {Yang},\ and\ \citenamefont {Vellekoop}}]{jang2017optical}%
  \BibitemOpen
  \bibfield  {author} {\bibinfo {author} {\bibfnamefont {M.}~\bibnamefont {Jang}}, \bibinfo {author} {\bibfnamefont {C.}~\bibnamefont {Yang}},\ and\ \bibinfo {author} {\bibfnamefont {I.}~\bibnamefont {Vellekoop}},\ }\bibfield  {title} {\bibinfo {title} {Optical phase conjugation with less than a photon per degree of freedom},\ }\href@noop {} {\bibfield  {journal} {\bibinfo  {journal} {Phys. Rev. Lett.}\ }\textbf {\bibinfo {volume} {118}},\ \bibinfo {pages} {093902} (\bibinfo {year} {2017})}\BibitemShut {NoStop}%
\bibitem [{\citenamefont {Wisal}\ \emph {et~al.}(2024{\natexlab{c}})\citenamefont {Wisal}, \citenamefont {Warren-Smith}, \citenamefont {Chen}, \citenamefont {Cao},\ and\ \citenamefont {Stone}}]{wisal2024theory_PRX}%
  \BibitemOpen
  \bibfield  {author} {\bibinfo {author} {\bibfnamefont {K.}~\bibnamefont {Wisal}}, \bibinfo {author} {\bibfnamefont {S.~C.}\ \bibnamefont {Warren-Smith}}, \bibinfo {author} {\bibfnamefont {C.-W.}\ \bibnamefont {Chen}}, \bibinfo {author} {\bibfnamefont {H.}~\bibnamefont {Cao}},\ and\ \bibinfo {author} {\bibfnamefont {A.~D.}\ \bibnamefont {Stone}},\ }\bibfield  {title} {\bibinfo {title} {Theory of stimulated brillouin scattering in fibers for highly multimode excitations},\ }\href@noop {} {\bibfield  {journal} {\bibinfo  {journal} {Phys. Rev. X}\ }\textbf {\bibinfo {volume} {14}},\ \bibinfo {pages} {031053} (\bibinfo {year} {2024}{\natexlab{c}})}\BibitemShut {NoStop}%
\bibitem [{\citenamefont {Boonzajer~Flaes}\ \emph {et~al.}(2018)\citenamefont {Boonzajer~Flaes}, \citenamefont {Stopka}, \citenamefont {Turtaev}, \citenamefont {De~Boer}, \citenamefont {Tyc},\ and\ \citenamefont {{\v{C}}i{\v{z}}m{\'a}r}}]{boonzajer2018robustness}%
  \BibitemOpen
  \bibfield  {author} {\bibinfo {author} {\bibfnamefont {D.~E.}\ \bibnamefont {Boonzajer~Flaes}}, \bibinfo {author} {\bibfnamefont {J.}~\bibnamefont {Stopka}}, \bibinfo {author} {\bibfnamefont {S.}~\bibnamefont {Turtaev}}, \bibinfo {author} {\bibfnamefont {J.~F.}\ \bibnamefont {De~Boer}}, \bibinfo {author} {\bibfnamefont {T.}~\bibnamefont {Tyc}},\ and\ \bibinfo {author} {\bibfnamefont {T.}~\bibnamefont {{\v{C}}i{\v{z}}m{\'a}r}},\ }\bibfield  {title} {\bibinfo {title} {Robustness of light-transport processes to bending deformations in graded-index multimode waveguides},\ }\href@noop {} {\bibfield  {journal} {\bibinfo  {journal} {Phys. Rev. Lett.}\ }\textbf {\bibinfo {volume} {120}},\ \bibinfo {pages} {233901} (\bibinfo {year} {2018})}\BibitemShut {NoStop}%
\end{thebibliography}%
	
\end{document}